\documentclass[sigplan,10pt]{acmart}

\usepackage[ruled,vlined,linesnumbered]{algorithm2e}
\usepackage{subcaption}
\usepackage{multirow}
\usepackage{latexml}
\SetKwInput{KwInput}{Input}
\SetKwInput{KwOutput}{Output}
\SetKwComment{tcp}{// }{}

\newcommand{\myparagraph}[1]{\noindent\textbf{#1}}
\DeclareMathOperator*{\argmax}{arg\,max}
\newcommand{\secref}[1]{\S\ref{#1}}
\newcommand{\subfigref}[2]{\iflatexml #2\else\subref{#1}\fi}

\newcommand{\rebuttalrev}[1]{{#1}}
\newcommand{\bugfixrev}[1]{{#1}}
\newcommand{\revised}[1]{{#1}}

\newenvironment{shepherd_2}{}{}

\AtBeginDocument{}

\copyrightyear{2026}
\acmYear{2026}
\setcopyright{cc}
\setcctype{by}
\acmConference[SIGOPS ATC 2026]{The 2026 ACM SIGOPS Annual Technical Conference}{November 15--18, 2026}{Hong Kong, China}
\acmBooktitle{The 2026 ACM SIGOPS Annual Technical Conference (SIGOPS ATC 2026), November 15--18, 2026, Hong Kong, China}
\acmDOI{10.1145/3828523.3857045}
\acmISBN{979-8-4007-2818-1/2026/11}

\makeatletter
\renewcommand{\@mkauthors}{\begingroup
  \gdef\@currentauthors{}%
  \global\setbox\mktitle@bx=\vbox{\unvbox\mktitle@bx\par\medskip
    \centering
    \noindent{\@authorfont
      Yunze Wei\textsuperscript{1,2}\enspace Kaicheng Yang\textsuperscript{2}\enspace Xiaohui Xie\textsuperscript{1}\enspace Dunkai Mao\textsuperscript{1}\enspace Peiying Tao\textsuperscript{2}\enspace Yikai Zhao\textsuperscript{2}\\
    Bolong Cui\textsuperscript{2}\enspace Chen Sun\textsuperscript{2}\footnotemark[1]\enspace Bingyang Liu\textsuperscript{2}\enspace Kun Tan\footnotemark[2]\enspace Kai Chen\textsuperscript{3}\enspace Yong Cui\textsuperscript{1}\footnotemark[1]}\par
    \smallskip
    \noindent{\@affiliationfont
      \textsuperscript{1}Tsinghua University\quad
      \textsuperscript{2}Huawei\quad
      \textsuperscript{3}Hong Kong University of Science and Technology}\par\bigskip}\endgroup}
\makeatother

\begin{document}

\title[D-Pingmesh]{D-Pingmesh: Dynamic Probing and Fault Localization for Reconfigurable Datacenter Networks}

\author{Yunze Wei}
\orcid{0009-0004-0618-512X}
\affiliation{\institution{Tsinghua University}\city{Beijing}\country{China}}
\affiliation{\institution{Huawei}\city{Beijing}\country{China}}
\author{Kaicheng Yang}
\orcid{0000-0001-6381-4026}
\affiliation{\institution{Huawei}\city{Beijing}\country{China}}
\author{Xiaohui Xie}
\orcid{0000-0001-9413-4461}
\affiliation{\institution{Tsinghua University}\city{Beijing}\country{China}}
\author{Dunkai Mao}
\orcid{0009-0000-4262-9446}
\affiliation{\institution{Tsinghua University}\city{Beijing}\country{China}}
\author{Peiying Tao}
\orcid{0009-0003-8286-9630}
\affiliation{\institution{Huawei}\city{Beijing}\country{China}}
\author{Yikai Zhao}
\orcid{0000-0003-2495-7774}
\affiliation{\institution{Huawei}\city{Beijing}\country{China}}
\author{Bolong Cui}
\orcid{0009-0006-5983-1492}
\affiliation{\institution{Huawei}\city{Beijing}\country{China}}
\author{Chen Sun}
\orcid{0000-0003-2480-2350}
\iflatexml\else\authornote{Co-corresponding authors: Chen Sun (sunchen48@huawei.com) and \\ Yong Cui (cuiyong@tsinghua.edu.cn).}\fi
\affiliation{\institution{Huawei}\city{Beijing}\country{China}}
\author{Bingyang Liu}
\orcid{0000-0002-4572-4956}
\affiliation{\institution{Huawei}\city{Beijing}\country{China}}
\author{Kun Tan}
\orcid{0000-0002-5876-9096}
\iflatexml\footnotemark[2]\else\authornote{Kun Tan is an independent researcher.}\fi
\affiliation{\institution{Unaffiliated}\city{Beijing}\country{China}}
\author{Kai Chen}
\orcid{0000-0003-2587-6028}
\affiliation{\institution{Hong Kong University of Science and Technology}\city{Hong Kong}\country{Hong Kong}}
\author{Yong Cui}
\orcid{0000-0002-5171-739X}
\iflatexml\footnotemark[1]\else\authornotemark[1]\fi
\affiliation{\institution{Tsinghua University}\city{Beijing}\country{China}}

\renewcommand{\shortauthors}{Yunze Wei et al.}

\begin{abstract}

Pingmesh monitors datacenter networks through periodic pings between
servers. In reconfigurable datacenter networks, topology engineering (TPE) and traffic
engineering (TE) make uniform server-pair probing inefficient: a pinglist wastes
budget retesting common paths while missing low-probability interfaces. Frequent control-plane updates further require rapid pinglist refresh, and cause path ambiguity at control-plane boundaries for fault localization.
We present D-Pingmesh, a probing and fault-localization system co-designed with
the TPE/TE control plane. Its control-plane-aware generator builds task-efficient pinglists by greedily admitting probes with high interface coverage under the current TPE/TE snapshot. An incremental replay-and-repair mechanism updates
pinglists across reconfigurations rather than recomputing from scratch. At
control-plane boundaries, a Rho-EM localizer jointly infers per-task
old/new-path mixture weights and the fault set through EM-style iteration.
We evaluate D-Pingmesh on an 8-pod reconfigurable topology under dynamic TPE/TE. 
\bugfixrev{At a 0.99 coverage target, D-Pingmesh uses 17.7$\times$ fewer tasks than Pingmesh and generates each pinglist about 5.1$\times$ faster than RD-Probe. }
In ns-3 simulation, boundary-aware localization improves micro F1
from 0.871 to 0.953 over the deTector PLL baseline. A testbed deployment
demonstrates second-scale fault localization with low host overhead.
These results demonstrate that D-Pingmesh is practical for reconfigurable
datacenter networks with control-plane-aware monitoring.
\end{abstract}



\maketitle
\pagestyle{plain}
\iflatexml
\footnotetext[1]{Co-corresponding authors: Chen Sun (sunchen48@huawei.com) and Yong Cui (cuiyong@tsinghua.edu.cn).}
\footnotetext[2]{Kun Tan is an independent researcher.}
\fi

\section{Introduction}

Network faults remain a critical source of availability incidents in modern
datacenters. Even when per-component failure rates appear small, they become
operationally significant at cluster scale: measurements of Google's TPUv4
deployment report average daily failure rates of 0.005\% for interconnect
cables and 0.04\% for optical circuit switches (OCSes)
~\cite{zu2024resiliency}. In clusters with thousands to tens of thousands of
devices, such rates make online network testing and accurate fault
localization necessary rather than optional.

Active probing remains one of the most practical mechanisms for this purpose.
By sending probe traffic between servers, it provides direct end-to-end
evidence about reachability and latency without requiring
pervasive switch instrumentation or changes to production applications.
Pingmesh established this model for large datacenter
networks~\cite{guo2015pingmesh}. Subsequent systems have improved probe
placement, coverage guarantees, service awareness, and fault localization for
different settings
~\cite{peng2017detector,ding2024rdprobe,liu2024rpingmesh,tan2019netbouncer,guo2025bytetracker}.
These systems, however, largely assume that the network state used to plan
probes remains stable within a measurement window: once a task is selected,
its observations can be associated with a small set of expected paths.

This assumption breaks down in modern datacenters, which increasingly treat the
network as a reconfigurable resource driven by two architectural control-plane
mechanisms. Topology engineering (TPE)
reconfigures the OCS to align connectivity with traffic
demand~\cite{poutievski2022jupiter,cao2023trod,ballani2020sirius,liu2023lightwave,shou2025infinitehbd,lei2026openoptics}.
Traffic engineering (TE) adjusts weighted forwarding splits over that topology
to accommodate shorter-term demand fluctuations~\cite{zhang2021gemini,liu2024figret,mao2026ssdo}. These mechanisms
can change which physical interfaces a given packet traverses at sub-minute to
minute timescales~\cite{mao2026ssdo}.
AI training and inference clusters further reinforce this trend: their recurring
communication patterns make TPE and TE especially effective at adapting the
network to demand~\cite{wang2023topoopt,khani2021sip}.

However, probing systems such as Pingmesh were designed and have
evolved independently of these control loops. They treat the network as a
static underlay and plan probes in terms of server pairs,
without consuming the current TPE/TE state as input.
Consider a TE update that changes the
forwarding weights between two pods: a probe task between the same two servers
can resolve to different interfaces before and after the update. A
monitor seeing only the server pair cannot determine whether a loss
observation belongs to the old path or the new one.
Across our measurement traces, 15--28\% of probe tasks change their resolved
path across a single TE update on average. The server pair, which traditional
probing uses as the stable unit of diagnosis, loses its anchoring power once TE
remaps the same pair to different interfaces.

\begin{shepherd_2}
A single timely detection may suffice for a persistent hard failure, but diagnosing gray packet loss requires repeated observations, and transient faults may disappear before the next stable window.
These observation requirements motivate continuous probing that gathers sufficient evidence while faults are present and correctly interprets observations across reconfigurations.
\end{shepherd_2}

\begin{shepherd_2}
Meeting these requirements calls for co-designing network monitoring with the TPE/TE control plane to overcome three challenges.
\end{shepherd_2}
First, TPE and TE induce
\emph{asymmetry} in interface hit probability. Under uniform server-pair selection, probes concentrate on high-probability interfaces and may starve
others in the same network state. Second, TPE/TE updates introduce \emph{dynamics}:
a pinglist optimized for one snapshot may lose coverage after the
next update, yet regenerating from scratch discards the
coverage contribution of tasks that remain valid.
Third, TPE/TE transitions create \emph{boundary inconsistency} for
diagnosis. Because TE updates deploy at pod granularity rather than
atomically, different parts of the network transiently
operate under different TE snapshots, so probe observations in the same
probe window may correspond to different path states.

To address these challenges, we present D-Pingmesh, a probing and
fault-localization system co-designed with the TPE/TE control plane. D-Pingmesh consumes current topology and
TPE/TE snapshots, and uses reconfiguration timing to schedule
probe windows at ping agents. For each candidate ping task, D-Pingmesh
determines the interfaces that the task can observe under the current
control-plane state and uses this information for both task planning and fault
interpretation.

D-Pingmesh tackles the asymmetry challenge by formulating probe generation
as an interface multicover problem under the current
control-plane state. It scores pod pairs by their expected coverage contribution, samples candidate five-tuples from high-scoring pairs,
and greedily admits tasks that advance the multicover objective. To handle TPE/TE dynamics,
D-Pingmesh maintains this multicover solution incrementally: it replays the
previous pinglist under the new snapshot, repairs residual coverage
deficits, and prunes redundant tasks without reducing achieved
coverage.

D-Pingmesh further addresses boundary inconsistency with Rho-EM, a fault
localizer that considers both pre- and post-update path hypotheses. For
aggregate probe results collected around a TE transition, Rho-EM assigns each
task a path-mixture weight $\rho_q$,
evaluates candidate faulty links through a likelihood model that estimates per-interface loss rates, covering both
complete link failures and gray packet-loss faults, and
refines the weights and fault set jointly via EM-style iteration.

\begin{figure}[t]
\centering
\begin{subfigure}[t]{0.65\linewidth}
  \centering
  \includegraphics[width=\linewidth]{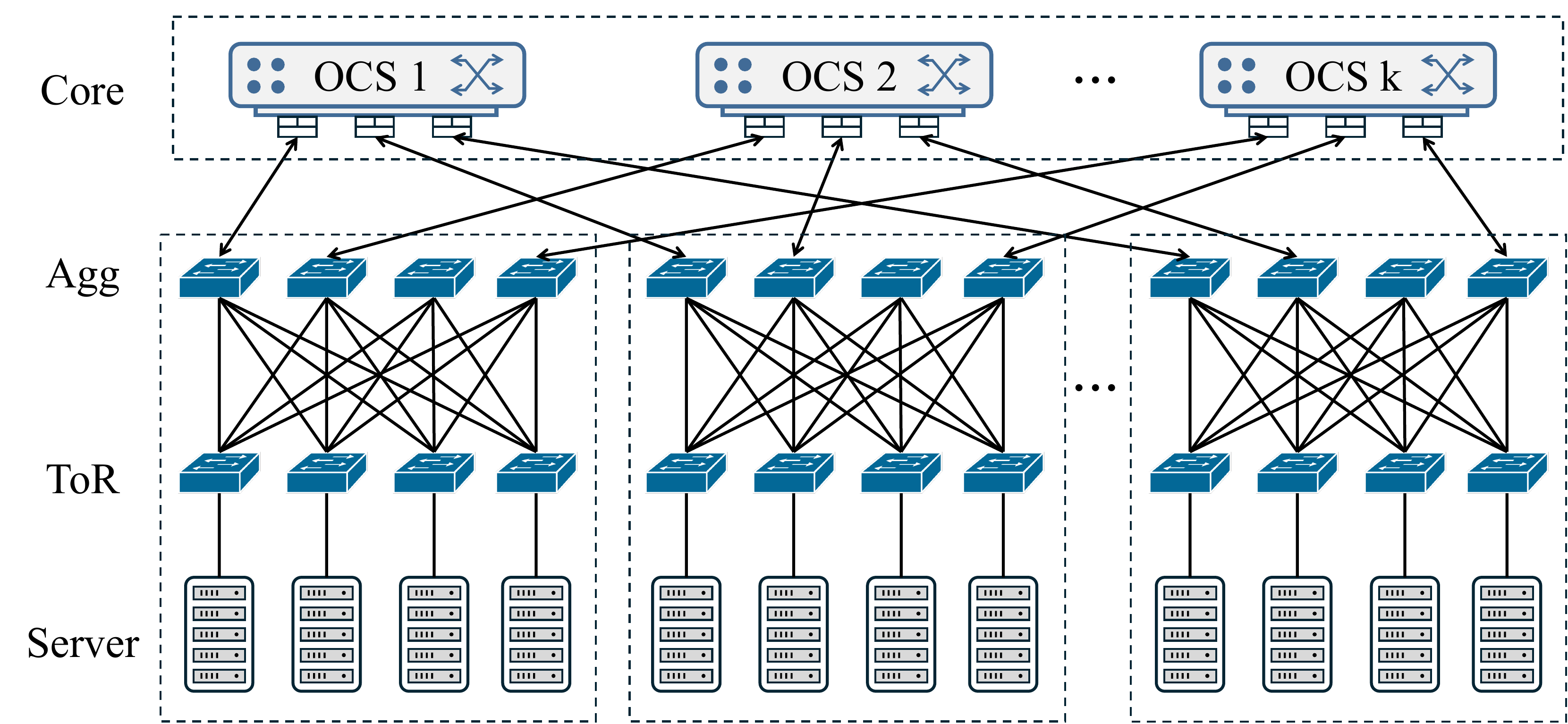}
  \caption{A reconfigurable datacenter network model with OCSes.}
  \label{fig:ocs-dcn-topology}
\end{subfigure}\hspace{0.06\linewidth}
\begin{subfigure}[t]{0.28\linewidth}
  \centering
  \includegraphics[width=\linewidth]{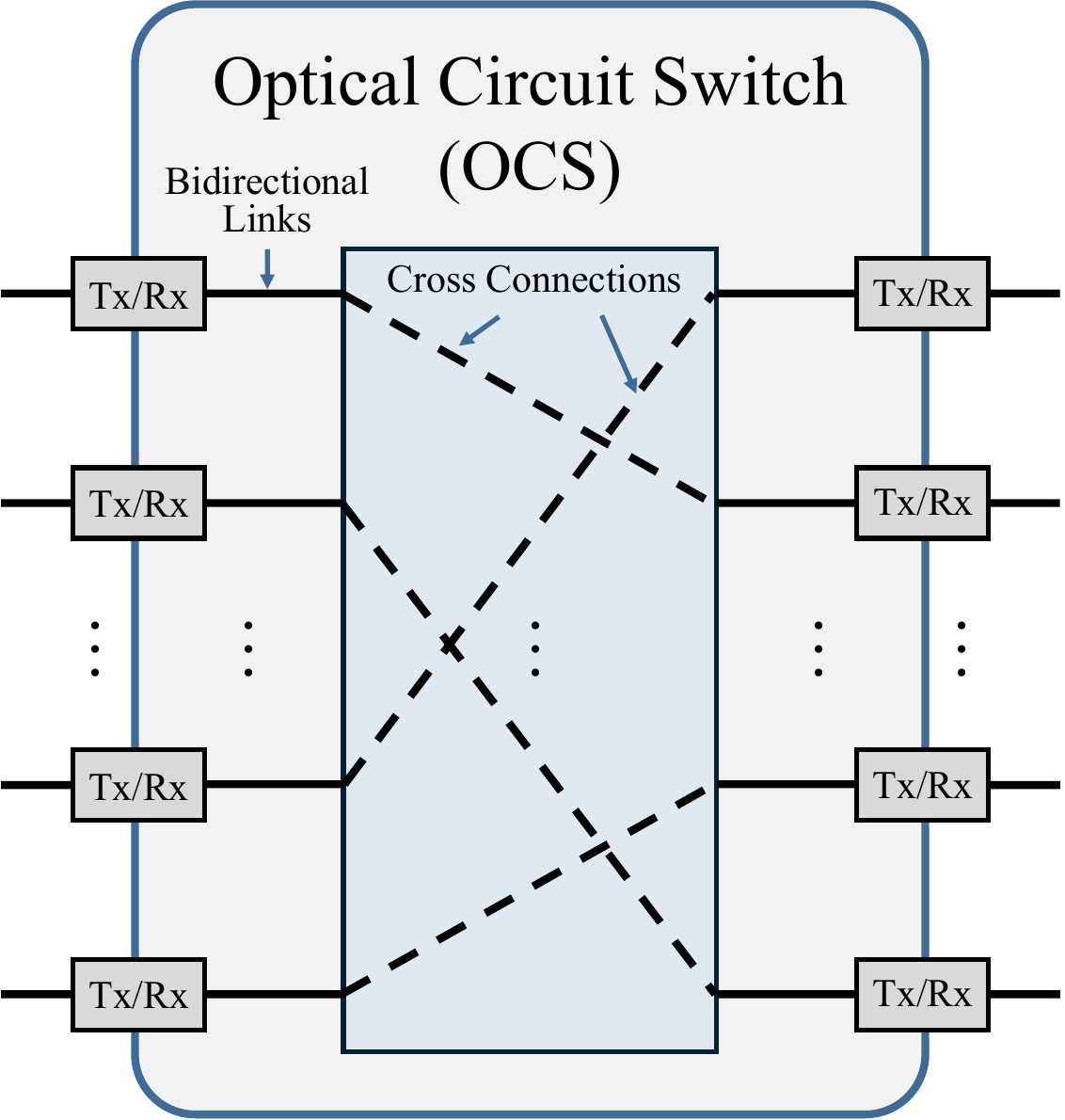}
  \caption{OCS abstraction.}
  \label{fig:ocs-abstraction}
\end{subfigure}
\caption{OCS-based reconfigurable datacenter network.}
\label{fig:OCS-DCN}
\end{figure}

\rebuttalrev{
We evaluate D-Pingmesh on an 8-pod, 2,048-server OCS-based topology, comparing it with Pingmesh, deTector, and RD-Probe. 
Our evaluation includes TE updates and controlled TPE transitions, with a separate scalability study extending to 64 pods and 16,384 servers.
}
At a 0.99 coverage target, D-Pingmesh reduces task count by
\bugfixrev{17.7$\times$ over Pingmesh and generation time by an estimated 5.1$\times$ over RD-Probe.}
In ns-3 simulation, Rho-EM improves F1 score of fault localization
from 0.871 with deTector PLL to 0.953 across varying fault counts and loss
rates. A testbed prototype further shows that D-Pingmesh localizes faults at second
scale with low host overhead.

We make the following contributions:

\begin{itemize}
\item We identify three monitoring challenges in reconfigurable datacenter
networks driven by TPE and TE: interface-hit asymmetry, coverage changes
across snapshots, and boundary inconsistency.

\item We design a control-plane-aware probe generation and incremental update
approach. It scores pod pairs by their expected coverage contribution
under the current TPE/TE snapshot, and incrementally repairs pinglists
across TPE/TE changes.

\item We design Rho-EM, a boundary-aware fault localizer that jointly refines
per-task old/new-path probabilities and a fault set via EM-style iteration,
supporting both link failures and gray packet-loss faults.

\item We evaluate D-Pingmesh under dynamic TPE/TE\linebreak inputs. It
achieves target coverage with fewer tasks and lower generation time
than Pingmesh, deTector, and RD-Probe. A testbed prototype
demonstrates second-scale fault localization with low host overhead.
\end{itemize}

\begin{table*}[t]
\centering
\caption{Comparison of monitoring systems from the perspective of
reconfigurable datacenter networks.} \label{tab:motivation-capability-gap}
\setlength{\tabcolsep}{8pt}
\begin{tabular}{p{0.41\linewidth}cccc}
\toprule
\textbf{Monitoring Requirement} & \emph{Pingmesh}~\cite{guo2015pingmesh} & \emph{deTector}~\cite{peng2017detector} & \emph{RD-Probe}~\cite{ding2024rdprobe} & \textbf{D-Pingmesh} \\
\midrule
Probe-efficient coverage under current TPE/TE state & $\times$ & $\triangle$ & $\triangle$ & \textbf{$\checkmark$} \\
Low-cost update after reconfiguration & $\times$ & $\times$ & $\triangle$ & \textbf{$\checkmark$} \\
Robust diagnosis at reconfiguration boundaries & $\times$ & $\times$ & $\times$ & \textbf{$\checkmark$} \\
\bottomrule
\end{tabular}

{\small\raggedleft $\checkmark$: supported, $\triangle$: partial support, $\times$: not
supported.\par}

\end{table*}

\section{Background and Motivation}
\label{sec:background-motivation}

\subsection{Reconfigurable Datacenter Networks}

We focus on reconfigurable datacenter networks that combine an electrical
intra-pod hierarchy with optical inter-pod connectivity, as sketched in
Figure~\ref{fig:OCS-DCN}(\subfigref{fig:ocs-dcn-topology}{a}). Within each pod, traffic is
forwarded through servers, Top-of-Rack switches (ToRs), and aggregation (Agg) switches. Across pods,
OCSes connect Agg uplinks to remote Agg ports,
replacing the fixed electrical spine of a conventional Clos network.
Such OCS-based designs appear in Google's Jupiter
Evolving~\cite{poutievski2022jupiter} and TROD~\cite{cao2023trod}.

Operating such a network requires two coupled control-plane decisions: TPE,
which establishes optical circuits, and TE,
which splits traffic over the resulting connectivity. Both are driven by
traffic demand to improve network utilization under changing
workloads~\cite{poutievski2022jupiter,zhang2021gemini,cao2023trod}. Measurements
from production datacenters confirm that inter-server traffic exhibits
significant locality and stability~\cite{roy2015inside}, supporting the
feasibility of demand-driven reconfiguration.

\rebuttalrev{
TPE changes inter-pod optical connectivity by programming the OCS's internal cross-connects, while the physical Agg-to-OCS uplinks remain fixed. 
These cross-connects pair selected OCS ports to form duplex optical circuits, as shown in Figure~\ref{fig:OCS-DCN}(\subfigref{fig:ocs-abstraction}{b})~\cite{poutievski2022jupiter}. 
}

TE determines how traffic is split over the resulting connectivity, installing
weighted next-hop groups at ToRs and Aggs. Recent TE systems compute these
weights to balance load and improve utilization under dynamic traffic
demand~\cite{zhang2021gemini,liu2024figret,mao2026ssdo}. TE can operate on
short timescales: recent work reports split-ratio adjustment periods from
10 seconds to 15 minutes~\cite{mao2026ssdo}. These weighted groups can be
realized via weighted-cost multipath (WCMP), where a switch hashes a five-tuple
(source/destination IP, ports, and protocol) to one member of the weighted group~\cite{zhou2014wcmp}. Consequently, two
packets between the same server pair may traverse different interfaces.

Direct OCS links match persistent traffic trends, but direct-only forwarding
cannot absorb bursty deviations. Existing designs therefore use TE to steer
selected traffic over \textit{one-hop detours} through intermediate
pods~\cite{poutievski2022jupiter,cao2023trod}. We call a path \textit{direct}
when it uses an active optical circuit toward the destination pod, and a
\textit{detour} when it passes through an intermediate pod before reaching
the destination. From a monitoring perspective, detours increase path
uncertainty: the same server pair may resolve to paths with different sets of
switch interfaces depending on whether the probe takes a direct or detour route.

Together, TPE and TE introduce within-snapshot asymmetry and across-snapshot
dynamics, challenging the uniform, stable forwarding assumptions of
conventional Pingmesh (Section~\ref{subsec:challenges}).

\subsection{Limitations of Prior Monitoring Systems}

Prior systems have made substantial progress in datacenter monitoring and fault
localization.
Pingmesh~\cite{guo2015pingmesh} demonstrates the feasibility of always-on,
server-based measurement at production scale, but its uniform server-pair probing
spends a large task budget to obtain broad visibility.
deTector~\cite{peng2017detector} reduces probing overhead by selecting a
topology-aware probe matrix that jointly considers coverage and loss
localization, but computing such matrices becomes expensive at large scale.
RD-Probe~\cite{ding2024rdprobe} further targets sufficient Layer-2 coverage
in complex production networks by combining randomized probing with
deterministic repair, but its randomized phase can still spend probes on
already well-covered components before the repair phase fills remaining gaps.
These systems are not designed around TPE and TE awareness,
limiting their ability to generate and update pinglists efficiently and to
localize faults accurately in dynamic reconfigurable networks.
Systems for explicit path control or path tracking, such as
NetBouncer~\cite{tan2019netbouncer} and ByteTracker~\cite{guo2025bytetracker},
are not directly applicable in OCS-based reconfigurable fabrics. We treat them
as complementary path-visibility mechanisms rather than direct baselines.

\rebuttalrev{
Production reports also illustrate the cost of increasing uniform probing.
Microsoft's IMC'18 study reports that Pingmesh limits full-mesh measurements to
servers within the same rack because of scaling concerns~\cite{roy2018cloud}.
RD-Probe reports that, in the production datacenter it studied, increasing a prior
Pingmesh-like system from three to four probe tasks per server pair would overload
some agents~\cite{ding2024rdprobe}.
}

Table~\ref{tab:motivation-capability-gap} compares prior systems along three
dimensions: probe-efficient coverage under current TPE/TE state, low-cost
pinglist update after reconfiguration, and robust diagnosis at reconfiguration
boundaries. \rebuttalrev{Repeated measurements are needed to diagnose gray packet
loss, while transient faults may disappear before the next stable window.}
These dimensions are coupled: \rebuttalrev{low-hit interfaces can accumulate
evidence slowly under repeated uniform probing,} a pinglist efficient for one
snapshot may lose coverage after TPE or TE changes, and observations near a
boundary may reflect paths from either side of the change.

\subsection{Challenges}\label{subsec:challenges}

\myparagraph{Challenge 1: Topology and Forwarding Asymmetry.}
Reconfigurable datacenter networks introduce both topology asymmetry and
forwarding asymmetry. Topology asymmetry comes from TPE, which determines which
cross-pod optical links are available in the current snapshot, so direct
connectivity can be uneven across pod pairs. Forwarding asymmetry comes from TE on top of this topology: TE installs weighted next-hop groups and may
split traffic between direct and one-hop detour paths. As a result, interface
hit probabilities can be highly skewed: some interfaces lie on
high-probability forwarding paths, while interfaces on low-weight next hops may be traversed by only a small fraction of candidate
probes. Endpoint-uniform probing then creates redundant observations on
high-probability interfaces, while low-probability interfaces may fail to reach
the target coverage count needed for diagnosis.

\begin{figure}[t]
\centering
\begin{subfigure}[t]{0.32\linewidth}
  \centering
  \includegraphics[width=\linewidth]{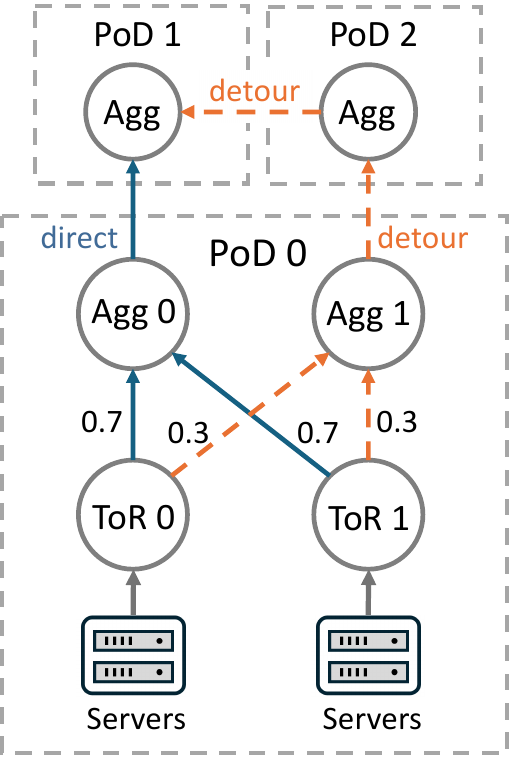}
  \caption{Example asymmetric forwarding state.}
  \label{fig:motivation-asymmetry-state}
\end{subfigure}\hspace{0.06\linewidth}
\begin{subfigure}[t]{0.38\linewidth}
  \centering
  \includegraphics[width=\linewidth]{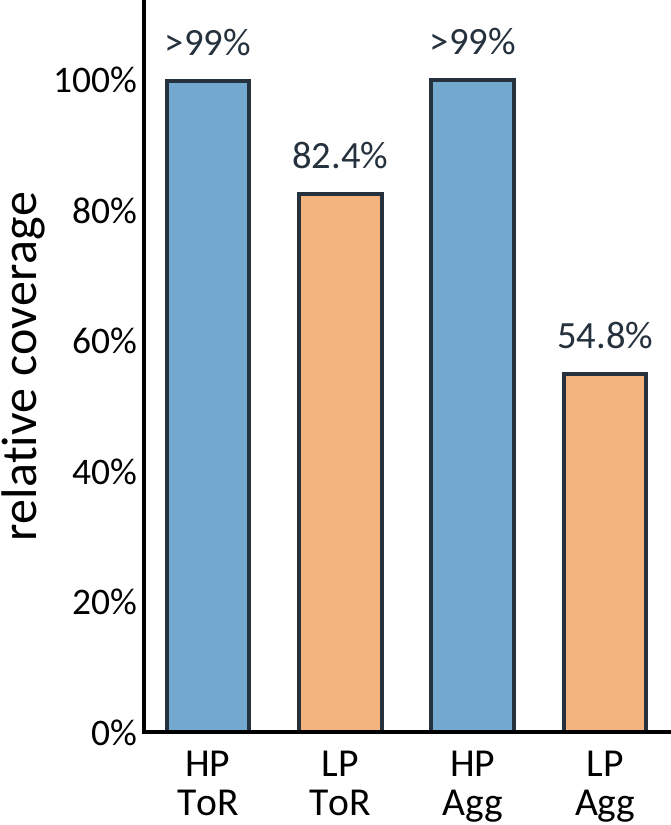}
  \caption{Uneven coverage under \rebuttalrev{Pingmesh-style} uniform probing.}
  \label{fig:motivation-asymmetry-coverage}
\end{subfigure}
\caption{Topology and forwarding asymmetry make uniform server-pair probing less
efficient.}
\Description{Two side-by-side panels: an asymmetric forwarding state with skewed direct and detour paths, and a coverage bar chart showing that low-probability interfaces remain harder to cover.}
\label{fig:motivation-asymmetry-efficiency}
\end{figure}

Figure~\ref{fig:motivation-asymmetry-efficiency} illustrates this effect with
a toy forwarding state. Traffic from pod 0 to pod 1 can either use Agg 0 to
reach pod 1 directly, or use Agg 1 to reach pod 2 and then take a one-hop
detour to pod 1. The ToR-to-Agg forwarding split is 7:3: Agg 0 leads to the
direct path, while Agg 1 leads to the one-hop detour path. For an interface with per-probe hit
probability $p$, Figure~\ref{fig:motivation-asymmetry-efficiency}(\subfigref{fig:motivation-asymmetry-coverage}{b}) shows the
probability of meeting a two-hit coverage target after 20 probes, i.e.,
$\Pr[X \geq 2]$ for $X \sim \mathrm{Binomial}(20,p)$. Under this target,
high-probability interfaces exceed 99\% coverage, while low-probability ToR
and Agg interfaces reach only 82.4\% and 54.8\% coverage, respectively. This
gap shows why blind uniform pinging is inefficient in the presence of topology
and forwarding asymmetry.

\begin{figure}[t]
\centering
\includegraphics[width=\linewidth]{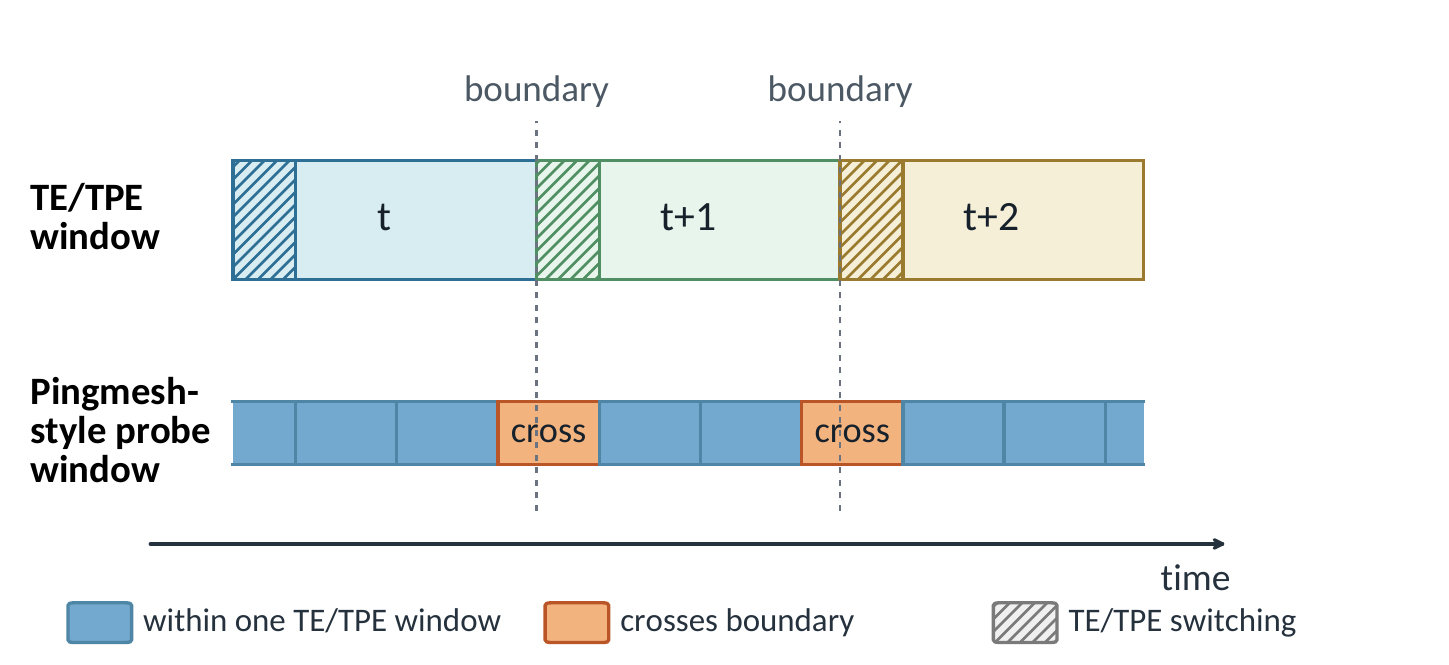}
\caption{Probe windows can overlap TE/TPE boundaries.}
\label{fig:motivation-boundary-windows}
\end{figure}

\myparagraph{Challenge 2: Topology and Forwarding Dynamics.}
After a TPE or TE update, a pinglist optimized for the prior snapshot may lose
coverage: the same five-tuple can resolve to different interfaces under the new
state. TE updates occur on timescales of 10 seconds to 15
minutes~\cite{mao2026ssdo}, so recomputing the full pinglist from scratch after
each change is expensive. Yet simply reusing the old pinglist leaves newly
exposed or remapped interfaces uncovered. The challenge is to update the
pinglist incrementally while preserving still-valid tasks, repairing coverage
deficits, and pruning redundancy with low overhead.

\begin{figure}[t]
\centering
\includegraphics[width=0.87\linewidth]{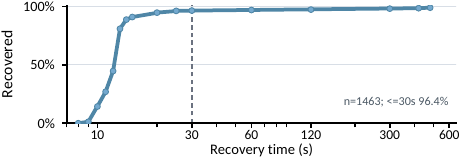}
\caption{Link-flap recovery times in production.}
\Description{A compact CDF of link-flap recovery times. The curve rises
quickly before 15 seconds and reaches 96.4 percent by 30 seconds.}
\label{fig:motivation-link-flap-recovery}
\end{figure}

\begin{figure}[t]
\centering
\begin{subfigure}[t]{0.4\linewidth}
  \centering
  \includegraphics[width=\linewidth]{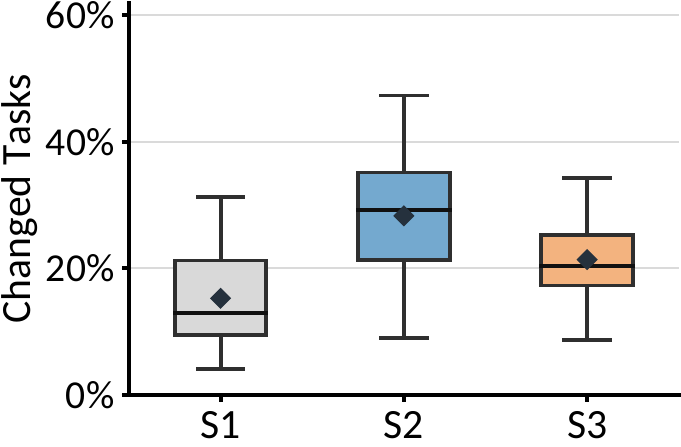}
  \caption{Path changes at TE boundaries.}
  \label{fig:motivation-boundary-change-ratio}
\end{subfigure}\hspace{0.06\linewidth}
\begin{subfigure}[t]{0.45\linewidth}
  \centering
  \includegraphics[width=\linewidth]{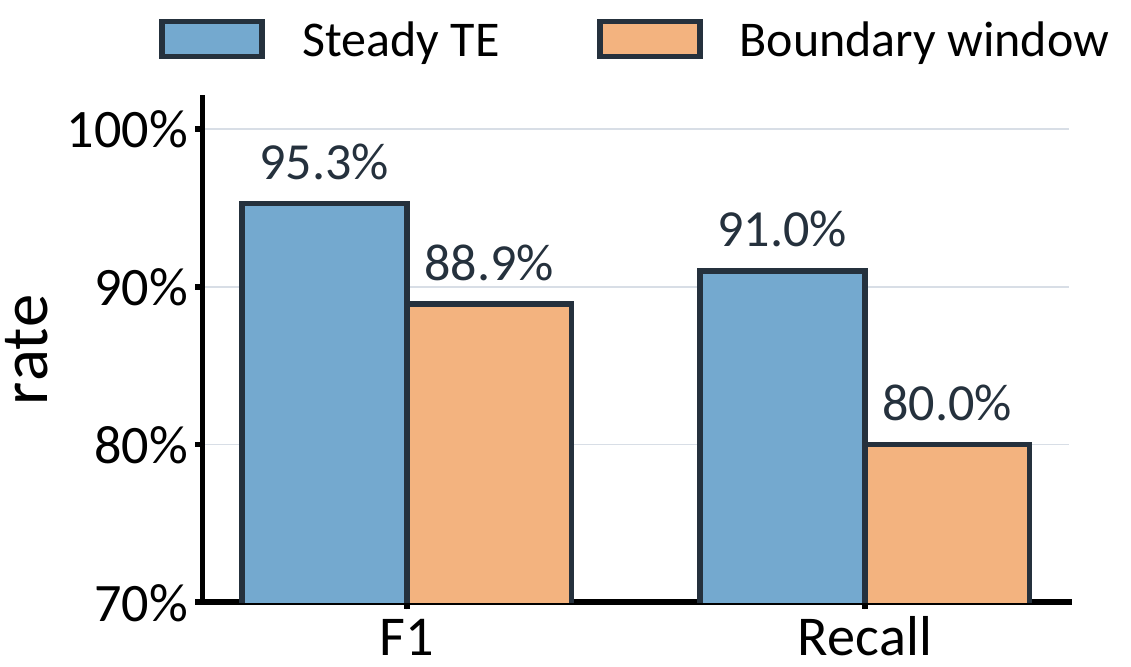}
  \caption{Fault localization degradation.}
  \label{fig:motivation-faultloc-boundary-gap}
\end{subfigure}
\caption{Boundary inconsistency affects probe paths and fault localization.}
\Description{Two side-by-side panels: a boxplot of changed-task ratios across three scenarios, and a bar chart comparing fault-localization accuracy and recall in steady and boundary conditions.}
\label{fig:motivation-boundary-evidence}
\end{figure}

\myparagraph{Challenge 3: Boundary Inconsistency.}
Traditional datacenter probing systems often operate under a comparatively
stable topology and forwarding model. In that setting, a probe degradation
signal can be interpreted against a relatively well-defined forwarding path.
Reconfigurable datacenter networks break this assumption. As shown in
Figure~\ref{fig:motivation-boundary-windows}, ping agents collect repeated
probe results over each probe window, such as 30 seconds, and report the
aggregate to the controller for analysis and fault localization, rather than
reporting each probe independently. Although a 30-second probe window is short,
it can occupy a significant fraction of a TE window when TE is updated within
15 minutes~\cite{mao2026ssdo}. Moreover, TE decisions may be
deployed at pod granularity rather than as a single global update, as in
Google's Jupiter Evolving~\cite{poutievski2022jupiter}. Around a TE boundary,
the resulting global observation is therefore a mixture of multiple local
forwarding states rather than a snapshot of one consistent network-wide state.
We refer to this mismatch as boundary inconsistency.

\rebuttalrev{
Boundary-window observations can be the only evidence of a transient fault. 
Among 1,463 link flaps recorded over 30 days in our production dataset, 96.4\% recovered within 30 seconds (Figure~\ref{fig:motivation-link-flap-recovery}), the example probe-window duration in Figure~\ref{fig:motivation-boundary-windows}. 
Waiting for a stable window can therefore miss the fault.
Continuous probing can retain and correctly interpret observations collected across the boundary.
}
Figure~\ref{fig:motivation-boundary-evidence} quantifies the effect: across TE
boundaries \revised{in three representative scenarios (S1--S3)}, 15.25--28.29\% of tasks change their resolved paths on average, and
fault localization F1 drops from 95.3\% to 88.9\% under a TE-boundary window
compared to steady state. These measurements require careful interpretation: if
analyzed only with the latest forwarding state, the localizer can assign loss
evidence to the wrong candidate interfaces.

\section{System Overview}
\label{sec:system-overview}

\begin{figure}[t]
\centering
\includegraphics[width=\linewidth]{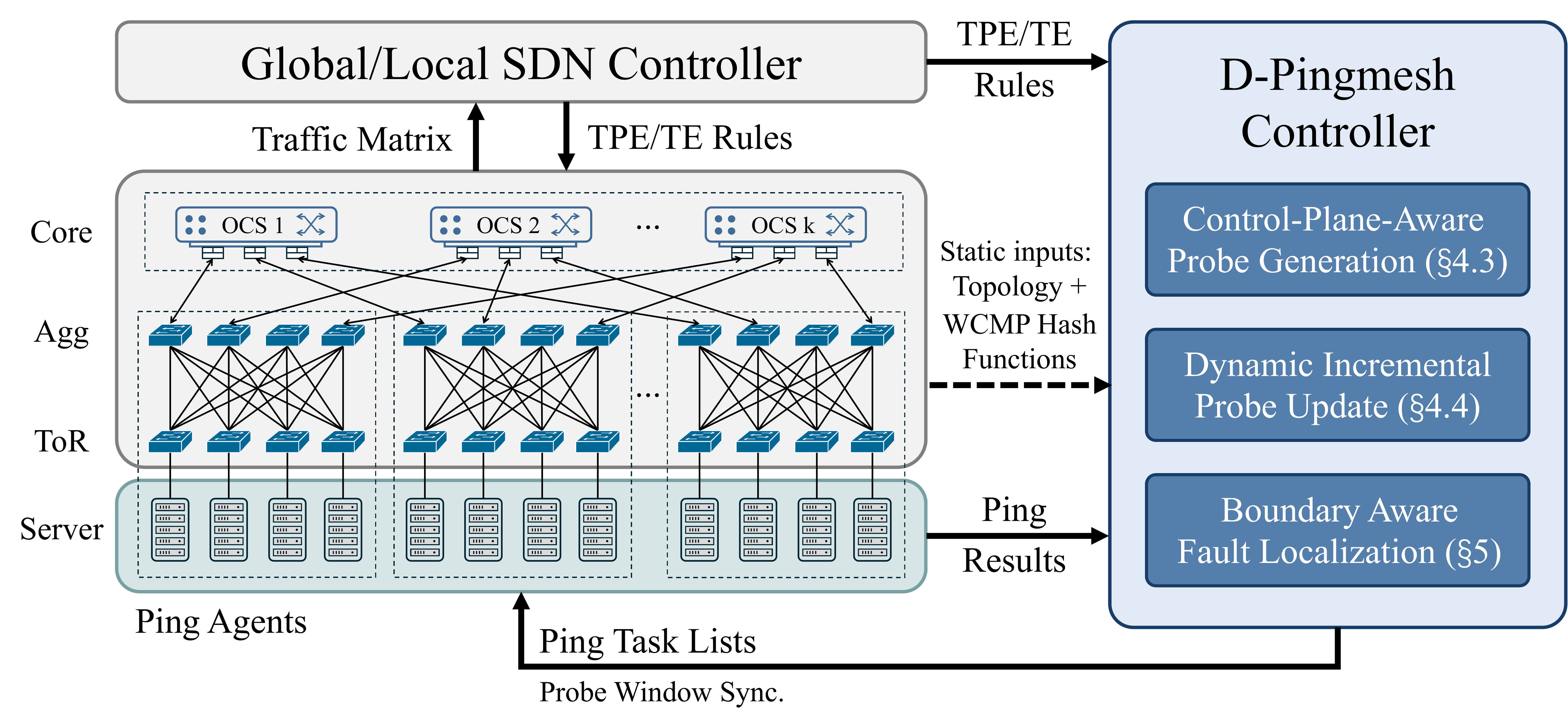}
\caption{System overview of D-Pingmesh.} \Description{System diagram showing a global or local SDN controller that
installs TPE and TE rules on an OCS-based datacenter network and shares those
rules with the D-Pingmesh controller. D-Pingmesh generates probe tasks,
synchronizes probe windows at ping agents, receives ping results, and performs
boundary-aware fault localization.}
\vspace{-1em}
\label{fig:system-overview}
\end{figure}

Figure~\ref{fig:system-overview} gives an overview of D-Pingmesh and its
control-plane-aware probing and diagnosis loop. The Software-Defined Networking (SDN) control plane
computes TPE and TE decisions from traffic demand and installs TPE
configurations in the optical circuit switched core and TE weighted next-hop
groups in the electrical switching hierarchy. D-Pingmesh runs alongside this
control plane and, for each reconfiguration slot, receives the current TPE
snapshot, TE forwarding state, and the schedule of upcoming TPE/TE updates.
It also receives static inputs, including the network topology and the WCMP
hash functions used by the switches.
With this control-plane awareness, the D-Pingmesh controller generates
slot-specific pinglists, aligns probe windows with TPE/TE updates,
and localizes faulty interfaces from aggregate probe results. 
\rebuttalrev{
Ping agents on host servers execute controller-computed pinglists in synchronized windows and return aggregate attempt and failure counts. 
Scheduling, incremental replay and repair, and localization reside at the controller; D-Pingmesh requires no switch modifications.
}

\myparagraph{Hash-aware path resolution.}
D-Pingmesh uses hash-aware path resolution to determine the expected forwarding
path of a candidate probe before the probe is executed. Commodity switch WCMP
hash functions can be modeled as deterministic header-to-next-hop
mappings~\cite{zhang2021hashing}. Given a concrete probe five-tuple and a
control-plane snapshot, the controller uses the hash functions to resolve the
next hop at each forwarding stage and reconstruct the complete expected path.
The generator uses this resolved path to decide whether a candidate task
improves interface coverage. The localizer uses the same resolver to associate
aggregate probe outcomes with candidate faulty interfaces.
\secref{sec:path-resolution} describes this path-resolution procedure.

\myparagraph{Control-plane-aware probe generation.} D-Pingmesh's probe
generator uses the current TPE and TE state to produce a probe-efficient task
list for each probe window. We formulate this as an interface multicover problem under skewed interface
hit probabilities (\secref{subsec:challenges}): each candidate five-tuple covers
its resolved interfaces, and the objective is to meet each target interface's
coverage requirement with as few tasks as possible.
D-Pingmesh uses a probability-guided algorithm that iteratively selects the
pod pair with the highest expected coverage gain, generates candidate tasks,
and uses hash-aware path resolution to accept only those that improve coverage.
\secref{subsec:formulation} and \secref{sec:asymmetry-aware-generation} present the
multicover formulation and the probability-guided generation algorithm.

\myparagraph{Dynamic incremental probe update.} TPE and TE updates can change
the resolved paths of existing probe tasks and the interfaces they cover.
Regenerating the full pinglist after every update wastes the coverage already
provided by tasks that remain useful under the new snapshot. D-Pingmesh
therefore maintains the pinglist incrementally: it re-resolves the previous
list under the new snapshot, repairs the remaining coverage deficits,
and removes redundancy from the merged list. The controller distributes the
updated pinglist to ping agents and synchronizes probe windows with the
TPE/TE update cycle.
\secref{sec:dynamic-incremental-update} details the incremental update
algorithm.

\myparagraph{Boundary-aware fault localization.}
To handle the boundary uncertainty introduced by frequent TPE and TE updates
(\secref{subsec:challenges}), we design Rho-EM, a boundary-aware fault
localization algorithm. Rho-EM operates on aggregate probe results from each
probe window. When a probe window contains a control-plane update
boundary, the same logical probe task may correspond to either the pre-update
path or the post-update path, while the aggregate result does not expose the
per-attempt path state. For each task, the algorithm resolves both path
hypotheses and maintains a per-task mixture over the old and new paths. It
scores candidate faulty interfaces using a likelihood over aggregate probe
counts and jointly refines the fault set and per-task path mixtures through
EM-style iteration.
\secref{sec:fault-localization} details the likelihood model and the Rho-EM
algorithm.

\section{Probe Generation and Update with Control-Plane Awareness}
\label{sec:probe-generation}

This section addresses the first two challenges from \secref{subsec:challenges}. 
D-Pingmesh first obtains the current TPE/TE snapshot
and uses hash-aware path resolution to map candidate five-tuples to their
link-level paths (\secref{sec:path-resolution}). The probe generator then selects
probe tasks that improve interface multicover under that snapshot
(\secref{sec:asymmetry-aware-generation}). To handle TPE/TE dynamics, we design
an incremental probe update algorithm that re-resolves the previous pinglist
under the new snapshot, repairs residual coverage deficits, and prunes
redundant tasks without reducing achieved coverage
(\secref{sec:dynamic-incremental-update}).

\subsection{Hash-Aware Path Resolution}
\label{sec:path-resolution}

D-Pingmesh uses hash-aware path resolution to interpret a concrete probe task
under a control-plane snapshot. Let $G$ denote the physical topology,
$\mathcal{T}_t$ the TPE snapshot at slot $t$, $\mathcal{E}_t$ the TE
snapshot at slot $t$, and $q$ a probe task with a concrete five-tuple. The
resolver returns
\[
R_t(q)=\mathrm{ResolvePath}(q,G,\mathcal{T}_t,\mathcal{E}_t),
\]
where $R_t(q)$ is the set of forward-path interfaces traversed by $q$ under
slot $t$.

$\mathrm{ResolvePath}$ combines offline hash awareness with the online TPE/TE
snapshot. The controller obtains the WCMP hash functions of ToR and Agg
switches. These can be modeled as deterministic header-to-next-hop
mappings~\cite{zhang2021hashing,fan2026bifrost}. Given the hash functions and
the current snapshot, $\mathrm{ResolvePath}$ uses the TE-installed WCMP groups
to resolve the egress port for $q$ at each ToR or Agg. At an Agg egress
connected to the OCS, the TPE snapshot maps the local port to the
remote pod and Agg reached by the configured optical circuit. Repeating these
steps reconstructs the complete path.

\rebuttalrev{
Probe generation, incremental update, and fault localization all use $R_t(q)$ under the assumption that the supplied TPE/TE state and switch hash behavior match active forwarding.
This consistency underpins interface coverage accounting and path-based fault attribution.
If actual forwarding differs from the resolved paths, D-Pingmesh may count untraversed interfaces as covered or attribute loss to the wrong interfaces.
Complementary runtime verification systems, including VeriDP and Hydra, check consistency
between specified and actual forwarding behavior~\cite{zhang2016mind,renganathan2023hydra}.
}

\subsection{Probe Task Generation Problem}\label{subsec:formulation}

The probe task generation problem is to choose a pinglist that covers each
target interface at least $k$ times under the current control-plane snapshot.
This $k$-coverage target provides multiple observations with overlapping paths
for each interface, giving fault localization the evidence needed to distinguish
candidate faulty interfaces.
For each slot $t$, the physical topology and the current TPE/TE snapshot induce
a path-resolved probe hypergraph: target interfaces are vertices, and each
feasible five-tuple $q$ contributes the hyperedge $R_t(q)$. We model probe task
generation as \emph{interface multicover} over this snapshot-specific
hypergraph.

Let $I_t$ be the reachable target interface set for slot $t$, and let $c_i$ be
the current coverage count for interface $i\in I_t$. D-Pingmesh uses an
end-to-end target set
$I_t=I_t^{\mathrm{up}}\cup I_t^{\mathrm{down}}$. For a task from pod $s$ to
pod $d$, $I_t^{\mathrm{up}}$ contains the forward interfaces from the source
server to the OCS, including server-to-ToR, ToR-to-Agg, and the Agg uplink
selected under the current TPE/TE snapshot. $I_t^{\mathrm{down}}$ contains the
forward interfaces from the destination Agg to the destination server,
including Agg-to-ToR and ToR-to-server links. For detour paths, the transit
Agg uplink is included in the upstream target set.
Only interfaces reachable by at least one bidirectionally valid candidate
under the current snapshot are included.

In the generation problem, interface $i$ is satisfied once its coverage counter
reaches $k$. Given the current counters $c_i$, the generator seeks a minimum
cardinality task set $Q$ that brings every reachable target interface to the
$k$-coverage target:
\begin{equation}
\label{eq:probe-generation-multicover}
\begin{aligned}
\min_{Q}\quad & |Q| \\
\text{s.t.}\quad &
c_i+\sum_{q\in Q}\mathbf{1}[i\in R_t(q)] \ge k,
\quad \forall i\in I_t .
\end{aligned}
\end{equation}
Here $\mathbf{1}[i\in R_t(q)]$ is one when interface $i$ lies on the
resolved path of task $q$ and zero otherwise.

Finding an exact minimum task set is NP-hard: even when $k=1$ and all candidate
hyperedges are pre-enumerated, the problem generalizes Set Cover. In our
setting, full enumeration is also impractical at datacenter scale because useful
hyperedges are revealed only after path resolution under the current TPE/TE
snapshot. D-Pingmesh therefore uses a probability-guided greedy approximation,
described next, to search for high-gain tasks without precomputing the full
candidate space.

\subsection{Asymmetry-Aware Probe Generation}
\label{sec:asymmetry-aware-generation}

\begin{algorithm}[!t]
\small
\caption{Asymmetry-Aware Probe Generation}
\label{alg:probe-generator}
\KwInput{snapshots $(G,\mathcal{T}_t,\mathcal{E}_t)$, target set $I_t$, initial counters $\{c_i\}_{i\in I_t}$, pair weights $W_{s,d}(i)$, target $k$, candidate budget $B$}
\KwOutput{accepted task list $Q$}
$Q\leftarrow \emptyset$\;
\ForEach{valid pod pair $(s,d)$}{
  $S(s,d)\leftarrow \sum_{i\in I_t: c_i<k} W_{s,d}(i)$\;
}
\While{$\exists i\in I_t$ with $c_i<k$}{
  $(s,d)\leftarrow \argmax_{(s',d')} S(s',d')$\;
  \If{$S(s,d)\le 0$}{\textbf{break}\;}
  $q^*\leftarrow\bot$; $g^*\leftarrow 0$\;
  \For{$b\leftarrow 1$ \KwTo $B$}{
    sample a candidate five-tuple $q$ for $(s,d)$\;
    $R_t(q)\leftarrow \mathrm{ResolvePath}(q,G,\mathcal{T}_t,\mathcal{E}_t)$\;
    \If{$R_t(q)=\emptyset$}{\textbf{continue}\;}
    $g\leftarrow \sum_{i\in R_t(q)\cap I_t} \mathbf{1}[c_i<k]$\;
    \If{$g>g^*$}{$q^*\leftarrow q$; $g^*\leftarrow g$\;}
  }
  \If{$q^*=\bot$}{
    $S(s,d)\leftarrow 0$\;
    \textbf{continue}\;
  }
  $Q\leftarrow Q\cup\{q^*\}$\;
  $c_i\leftarrow c_i+1$ for each $i\in R_t(q^*)\cap I_t$\;
  update $S(s',d')$ for pairs affected by newly satisfied interfaces\;
}
\Return{$Q$}\;
\end{algorithm}

Algorithm~\ref{alg:probe-generator} presents D-Pingmesh's
probability-guided greedy generator for approximating the multicover objective
in \eqref{eq:probe-generation-multicover}. The generator repeatedly scores
source-destination pod pairs by their expected contribution to unsatisfied
interfaces, resolves candidate five-tuples for the selected pair, and accepts
the probe task with the largest resolved marginal coverage.

\myparagraph{Interface hit probabilities.}
The pod pair scores in Algorithm~\ref{alg:probe-generator} are built from
interface hit probabilities. For each target interface $i$ and ordered pod pair
$(s,d)$, D-Pingmesh computes a pair-specific hit probability $W_{s,d}(i)$ as
the probability that probes from $s$ to $d$ expose $i$ under the current
TPE/TE snapshot.
Let $\mathrm{TEWeight}(x,a,d)$ denote the normalized forwarding probability
assigned by the current TE snapshot to egress interface $a$ at switch $x$ for
traffic toward destination pod $d$.
For both ToR uplinks and Agg uplinks, the hit probability equals the
probability of reaching the switch multiplied by the corresponding
$\mathrm{TEWeight}$ term.

For a ToR uplink interface $i=(v,a)$ on ToR switch $v$ in source pod $s$,
let $P_s(v)=\Pr[\text{source ToR}=v\mid s]$ denote the probability that the
source server selected from pod $s$ attaches to ToR $v$. Then
\[
W_{s,d}(i) = P_s(v)\cdot \mathrm{TEWeight}(v,a,d).
\]
Let $\mathrm{ToR}(s)$ be the ToR switches in pod $s$. When servers are
selected uniformly in a homogeneous pod, $P_s(v)$ simplifies to
$1/|\mathrm{ToR}(s)|$.

For an Agg uplink interface $i=(u,a)$ from Agg switch $u$ to the OCS, the
switch arrival probability is determined by the ToR-to-Agg choices in pod $s$.
Let $\pi_u(s,d)$ be the probability of reaching $u$:
\[
\pi_u(s,d)=
  \sum_{v\in \mathrm{ToR}(s)}
  P_s(v)\cdot \mathrm{TEWeight}(v, v\rightarrow u, d).
\]
Multiplying the arrival probability by the local TE weight gives
\[
W_{s,d}(i)=\pi_u(s,d)\cdot \mathrm{TEWeight}(u,a,d).
\]

For destination side downlinks, once a probe reaches the destination pod's Agg
switch, the remaining path to the target server follows a single ToR-to-server
link determined by the server's rack placement, with no additional TE choices.
When destination servers are sampled uniformly, the hit probability is spread
uniformly across the corresponding downlink interfaces.
The hit probabilities are recomputed once per control-plane snapshot.

\myparagraph{Pod pair scoring.}
Using the hit probabilities defined above, the generator scores each ordered
pod pair $(s,d)$ by its expected contribution to interfaces that remain below
the $k$-coverage target:
\[
S(s,d) = \sum_{i\in I_t: c_i<k} W_{s,d}(i).
\]
$S(s,d)$ estimates the expected number of unsatisfied interfaces that a task
from $s$ to $d$ would cover. The generator uses these scores to prioritize pod
pairs likely to advance the multicover objective. The scores serve as a ranking
heuristic: after selecting a pod pair, D-Pingmesh samples candidate tasks
within that pair, resolves their paths, and credits coverage according to
$R_t(q)$.

Because $S(s,d)$ changes as interfaces reach $k$, D-Pingmesh maintains pod
pair scores in a max-priority queue. Pair-to-interface and interface-to-pair
indexes identify scores affected by newly satisfied interfaces, and outdated
entries are recomputed lazily at the top of the queue.

The candidate budget $B$ controls how many candidate tasks are sampled for the
selected pod pair. D-Pingmesh samples up to $B$ candidates, resolves each, and
accepts the one with the largest marginal gain under the current counters.
This budget balances coverage quality against generation overhead: sampling a
single task can miss high-gain candidates, while sampling many increases
resolution cost. The search concentrates on pod pairs with high expected
coverage gain. We conducted a sensitivity analysis and found that budgets
above $B=3$ offer only modest additional task reduction at higher per-task
computation cost; we therefore use $B=3$ by default.

\subsection{Dynamic Incremental Probe Update}
\label{sec:dynamic-incremental-update}

A straightforward approach is to discard the existing pinglist after each
TPE/TE reconfiguration and invoke Algorithm~\ref{alg:probe-generator} from
empty counters. This cold-start approach is wasteful, since many tasks from the
prior pinglist typically remain useful under the new snapshot. D-Pingmesh
instead formulates probe update as \emph{dynamic multicover maintenance}: given
a base pinglist $Q_{\mathrm{base}}$ and the current snapshot
$(\mathcal{T}_t,\mathcal{E}_t)$, the objective is to produce an updated
pinglist $Q_t$ that reuses useful tasks from $Q_{\mathrm{base}}$, fills
residual coverage deficits, and removes redundancies through
coverage-preserving pruning.
Algorithm~\ref{alg:incremental-update} accomplishes this in three
phases: re-resolution, repair, and pruning.

\begin{algorithm}[!t]
\small
\caption{Dynamic Incremental Probe Update}
\label{alg:incremental-update}
\KwInput{base pinglist $Q_{\mathrm{base}}$, snapshots $(G,\mathcal{T}_t,\mathcal{E}_t)$, target set $I_t$, target $k$, pair weights $W_{s,d}(i)$, candidate budget $B$}
\KwOutput{updated pinglist $Q_t$}
\tcc{Phase 1: Re-resolution}
$Q_{\mathrm{keep}}\leftarrow \emptyset$; $c_i\leftarrow 0$ for all $i\in I_t$\;
\ForEach{$q\in Q_{\mathrm{base}}$}{
  $R_t(q)\leftarrow \mathrm{ResolvePath}(q,G,\mathcal{T}_t,\mathcal{E}_t)$\;
  \If{$R_t(q)\cap I_t\neq\emptyset$}{
    $Q_{\mathrm{keep}}\leftarrow Q_{\mathrm{keep}}\cup\{q\}$\;
    $c_i\leftarrow c_i+1$ for each $i\in R_t(q)\cap I_t$\;
  }
}
\tcc{Phase 2: Repair}
$Q_{\mathrm{repair}}\leftarrow \textsc{ProbeGenerate}(G,\mathcal{T}_t,\mathcal{E}_t,I_t,\{c_i\},W_{s,d},k,B)$\;
\tcc{Phase 3: Pruning}
$Q_{\mathrm{all}}\leftarrow$ deduplicate $Q_{\mathrm{keep}}\cup Q_{\mathrm{repair}}$\;
$\hat c_i\leftarrow \min\{\,k,\ |\{q\in Q_{\mathrm{all}}: i\in R_t(q)\}|\,\}$ for all $i\in I_t$\;
$Q_t\leftarrow \emptyset$; $c'_i\leftarrow 0$ for all $i\in I_t$\;
\While{$\exists i\in I_t$ with $c'_i<\hat c_i$}{
  $q^*\leftarrow \argmax_{q\in Q_{\mathrm{all}}\setminus Q_t}
    \sum_{i:\,c'_i<\hat c_i} \mathbf{1}[i\in R_t(q)]$\;
  \If{$\sum_{i:\,c'_i<\hat c_i} \mathbf{1}[i\in R_t(q^*)] = 0$}{\textbf{break}\;}
  $Q_t\leftarrow Q_t\cup\{q^*\}$\;
  $c'_i\leftarrow c'_i+1$ for each $i\in R_t(q^*)\cap I_t$\;
}
\If{$\exists i\in I_t$ with $c'_i<\hat c_i$}{$Q_t\leftarrow Q_{\mathrm{all}}$\;}
\Return{$Q_t$}\;
\end{algorithm}

\myparagraph{Re-resolution.}
In the re-resolution phase (Phase~1 of
Algorithm~\ref{alg:incremental-update}), each task $q\in Q_{\mathrm{base}}$ is
re-resolved against the current TPE/TE snapshot\rebuttalrev{, after the target
interface set $I_t$ has been recomputed for that snapshot}.
Tasks whose re-resolved paths still intersect $I_t$ are retained in
$Q_{\mathrm{keep}}$, with their covered interfaces credited to counters $c_i$.
Tasks whose paths no longer intersect any target interface are discarded.
The resulting counters capture the residual coverage of the prior pinglist
under the new snapshot.

\myparagraph{Repair.}
The repair phase (Phase~2) invokes
\textsc{ProbeGenerate} (Algorithm~\ref{alg:probe-generator}) with the
inherited counters. Since the generator scores pod pairs only over interfaces
with $c_i<k$, repair focuses on filling the residual deficits,
producing $Q_{\mathrm{repair}}$ that complements $Q_{\mathrm{keep}}$.

\myparagraph{Pruning.}
The merged set $Q_{\mathrm{all}} = Q_{\mathrm{keep}} \cup Q_{\mathrm{repair}}$
may contain redundant tasks, since the two phases independently cover
overlapping interfaces and some interfaces may exceed $k$ coverage.
The pruning phase (Phase~3) eliminates this redundancy while preserving the
achieved coverage level for every interface.

For each interface $i\in I_t$, let $n_i=|\{q\in Q_{\mathrm{all}}: i\in R_t(q)\}|$
be the number of tasks in the merged set that cover $i$. The pruning target is
$\hat c_i = \min(k, n_i)$, which caps the preserved coverage at $k$.
D-Pingmesh then greedily selects a
subset $Q_t\subseteq Q_{\mathrm{all}}$ to meet these targets, as shown in
Phase~3 of Algorithm~\ref{alg:incremental-update}.
If the greedy selection terminates before all targets are met,
D-Pingmesh defaults to $Q_{\mathrm{all}}$, guaranteeing that pruning never
reduces the coverage achieved by re-resolution and repair.

\section{Fault Localization Across Reconfiguration Boundaries}
\label{sec:fault-localization}

Probe generation selects which tasks to measure, while fault localization
inverts the resulting aggregate observations into a ranked list of likely
faulty interfaces. Traditional datacenter networks map each probe to a
deterministic path, but reconfigurable networks break this assumption: around
a TPE or TE update, the same task can traverse different paths, and an
observation near a reconfiguration boundary could belong to either the pre-
or post-update state. This ambiguity arises because TE updates deploy at pod
granularity rather than atomically across the network.

D-Pingmesh addresses this with Rho-EM, a boundary-aware localizer
that maintains both pre- and post-update path hypotheses for each task. Rho-EM
treats the per-attempt path as a hidden variable and jointly refines per-task
mixture weights and the fault set through EM-style iteration. It consumes only
aggregate success and failure counts per task within each probe window
together with the corresponding control-plane snapshots, and returns a
predicted fault set with a ranked suspicion list.

\subsection{Observation Model}

Let $Q$ be the set of probe tasks in one probe window. Each task
$q\in Q$ carries a concrete five-tuple and an aggregate observation
$y_q=(n_q,k_q)$, where $n_q$ is the number of attempts and $k_q$ is the number
of failed or timed-out attempts within the probe window. In our setting, each task is probed once per second over a 30-second probe window, so the aggregate observation summarizes up to 30 attempts whose paths can differ around a boundary.

For each task, D-Pingmesh resolves two path hypotheses $R_q^{o}$ and $R_q^{n}$
under the pre- and post-update snapshots using the hash-aware path resolver from
\secref{sec:path-resolution}.
\rebuttalrev{
For a changed task, Rho-EM maintains a mixture weight $\rho_q\in[0,1]$, the probability that a given attempt followed the complete post-update path.
The aggregate observation can mix attempts on these two paths.
An individual attempt traversing a hybrid path with old and new forwarding states at different hops falls outside this model.
}

\rebuttalrev{Let $E$ be the candidate faulty interface set comprising monitored
interfaces traversed by probe tasks under either the pre- or post-update
forwarding state. This union retains interfaces present in only one of the
two states' task paths.} For a candidate fault set
$F\subseteq E$, let $\theta_e$ be the estimated loss rate of interface $e$. The
failure probability of a task path $R$ is modeled by interface survivals:
\begin{equation*}
p_q(F,R)
= 1 - \prod_{e\in F\cap R}(1-\theta_e).
\end{equation*}
Each faulty interface on the path independently contributes to the failure
probability, so the model naturally handles paths that traverse multiple faults.

For a changed task, the failure probability under the mixture is
$p_q(F;\rho_q)=(1-\rho_q)p_q(F,R_q^{o})+\rho_q p_q(F,R_q^{n})$.
The log-likelihood of the observed counts is
\begin{equation}
\ell_q(F;\rho_q)
= k_q\log p_q(F;\rho_q)
  +(n_q-k_q)\log\bigl(1-p_q(F;\rho_q)\bigr).
\label{eq:mixed-loglik}
\end{equation}

Successful attempts weaken the evidence for interfaces on paths that largely
succeeded, while failures strengthen the evidence for interfaces on the old or
new paths.

\begin{algorithm}[t]
\small
\caption{Rho-EM Fault Localization}
\label{alg:rho-em-localization}
\KwInput{aggregate task records $Q$, snapshots $(G,\mathcal{T}_{o},\mathcal{E}_{o})$ and $(G,\mathcal{T}_{n},\mathcal{E}_{n})$, candidate interfaces $E$ \rebuttalrev{traversed by probe tasks under either snapshot}, initial mixture $\rho_0$, iteration cap $M$}
\KwOutput{predicted fault set $F$ and ranked suspicion list}
\tcc{Phase 1: Boundary path hypotheses}
\ForEach{$q\in Q$}{
  $R_q^{o}\leftarrow\mathrm{ResolvePath}(q,G,\mathcal{T}_{o},\mathcal{E}_{o})$\;
  $R_q^{n}\leftarrow\mathrm{ResolvePath}(q,G,\mathcal{T}_{n},\mathcal{E}_{n})$\;
  $\rho_q\leftarrow\rho_0$ if $R_q^{o}\neq R_q^{n}$, otherwise $\rho_q\leftarrow 0$\;
}
\tcc{Phase 2: Greedy selection and rho refinement}
\For{$r\leftarrow 0$ \KwTo $M$}{
  $(F,\Theta)\leftarrow \textsc{GreedySelect}(Q,E,\{R_q^{o},R_q^{n},\rho_q\})$\;
  compute ranked suspicion scores from the marginal gains\;
  \If{$r=M$}{\textbf{break}\;}
  \ForEach{changed task $q$}{
    $\rho_q\leftarrow\textsc{UpdateRho}(n_q,k_q,R_q^{o},R_q^{n},F,\Theta)$\;
  }
  \If{$\max_q|\Delta\rho_q|<\tau$}{\textbf{break}\;}
}
\Return{$F$ and all candidates ranked by suspicion score}\;
\end{algorithm}

\subsection{Rho-EM Fault Localization Algorithm} 

Rho-EM jointly refines the fault candidates and per-task path mixture weights
$\rho_q$ through an EM-style coordinate procedure~\cite{dempster1977maximum}.
In each iteration, Rho-EM first selects a small fault set $F$ that improves the
aggregate log-likelihood under the current $\rho_q$, then updates each changed
task's $\rho_q$ toward the hypothesis that better explains its counts under the
selected $F$. Algorithm~\ref{alg:rho-em-localization} outlines the full
procedure; the two key subroutines are detailed below.

\myparagraph{Greedy fault selection.}
The \textsc{GreedySelect} subroutine starts from an empty
fault set $F=\emptyset$ and iteratively adds
the candidate interface that yields the largest marginal log-likelihood gain.
For a candidate $e\notin F$, the marginal gain of adding $e$ to the
current fault set is
\begin{equation*}
\Delta(e\mid F)=
\sum_{q\in Q}\ell_q(F\cup\{e\};\rho_q)
-
\sum_{q\in Q}\ell_q(F;\rho_q).
\end{equation*}
Selection stops when the best gain falls below a threshold or a maximum fault
count is reached.

\myparagraph{Rho update.}
After selection, the \textsc{UpdateRho} subroutine recomputes each changed
task's old- and new-path failure probabilities under the fixed fault set
$\widehat{F}$ and its estimated loss rates, denoted
$p_q^{o}=p_q(\widehat{F},R_q^{o})$ and
$p_q^{n}=p_q(\widehat{F},R_q^{n})$. Rho-EM then chooses $\rho_q$ as the
maximum a posteriori (MAP) estimate:
the first line is the log-likelihood from Eq.~(\ref{eq:mixed-loglik}) with
$\rho$ as the variable, and the second line is a beta prior that prevents
degenerate estimates when evidence is weak.
\begin{equation*}
\begin{aligned}
\rho_q^{\star}
=\arg\max_{\rho\in[\rho_{\min},\rho_{\max}]}
&\; k_q\log p_q(\rho)
 +(n_q-k_q)\log\bigl(1-p_q(\rho)\bigr) \\
 &+(\alpha-1)\log\rho+(\beta-1)\log(1-\rho).
\end{aligned}
\end{equation*}
where $p_q(\rho)=(1-\rho)p_q^{o}+\rho p_q^{n}$. The update is damped by
blending with the previous $\rho_q$ value for numerical stability.

Iterations stop when the maximum rho change drops below a threshold or a small
cap $M$ is reached. With $M=0$, only the initial rho-weighted selection runs,
giving the fixed-mixture fallback used in our iteration ablation
(\secref{subsec:fault-loc}).

\begin{figure*}[!t]
\centering
\includegraphics[width=0.96\textwidth]{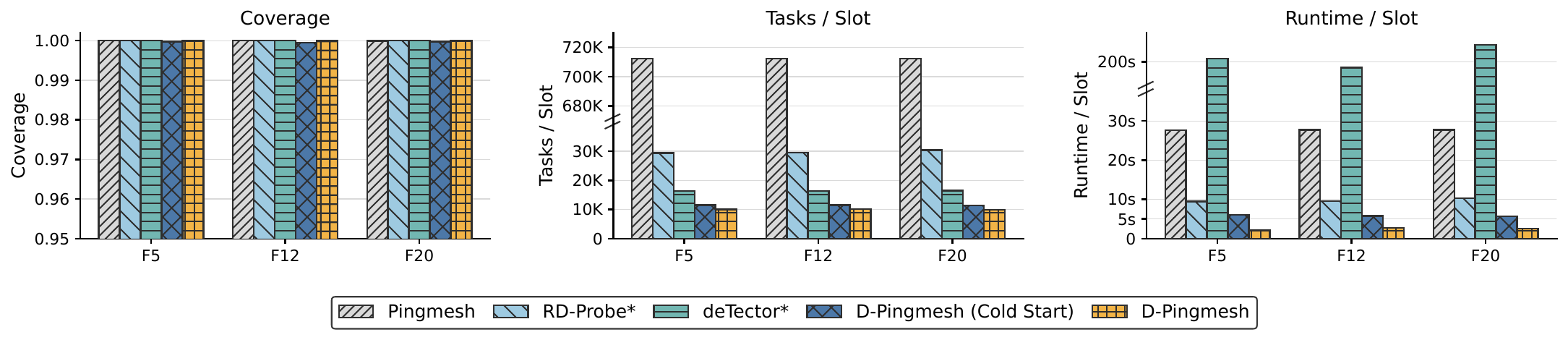}
\caption{Probe generation comparison under three representative traces (F5, F12,
and F20) across 96 slots.}
\Description{Three grouped bar charts compare Pingmesh, deTector*, RD-Probe*,
D-Pingmesh (Cold Start), and D-Pingmesh on coverage, tasks per slot, and
runtime per slot for F5, F12, and F20.}
\label{fig:main-five-algo}
\end{figure*}

\section{Evaluation}
\label{sec:evaluation}

We evaluate D-Pingmesh's probe-generation efficiency
(\S\ref{subsec:generator}), localization accuracy across TE reconfiguration
boundaries (\S\ref{subsec:fault-loc}), and practicality on a physical testbed
(\S\ref{subsec:physical-testbed}).

\subsection{Experimental Setup}

\myparagraph{Topology and traces.} The main experiments use an 8-pod reconfigurable
network. Each pod has 8 ToR switches and 8 Agg switches, both radix 64 with
32 downlinks and 32 uplinks, and 32 servers per ToR, for 2,048 servers in total.
16 OCSes with 128 100~Gbps ports each interconnect the pods; OCSes are grouped
by Agg index, so that Agg~$i$ of every pod is wired to the same pair of OCSes.
In total, the network comprises 2,192 nodes (2,048 servers, 64 ToR switches,
64 Agg switches, and 16 OCSes) and 12,288 interfaces.

\rebuttalrev{We synthesize pod-to-pod traffic matrices from the 22 production traffic-skew CDFs reported in Figure~5 of Gemini~\cite{zhang2021gemini}.}
\rebuttalrev{Each trace spans one day at one-second granularity, with average and peak offered loads of 30\% and 70\% of network capacity, and is aggregated into 96 fifteen-minute TE windows.}
For each trace, we run Gemini to jointly optimize topology and routing, producing the day-level TPE snapshot and 96 per-window TE snapshots that drive all probe generation algorithms.
The main experiments use three CDFs spanning a range of inter-pod traffic skew: F5, F12, and F20.
\rebuttalrev{Within-trace transition experiments evaluate TE changes under fixed TPE. 
We separately test incremental generation across controlled TPE transitions, which change OCS cross-connects together with their associated TE state.}

\myparagraph{Algorithms and baselines.} We compare five probe generation algorithms,
all targeting $k=4$ coverage per interface per slot:
(1)~\emph{Pingmesh}, the conventional server-pair-oriented probing
baseline~\cite{guo2015pingmesh}; (2)~\emph{deTector*}, which adapts
deTector~\cite{peng2017detector} by applying the PMC component to resolved
inter-switch paths and adding a server--ToR supplement of local same-ToR server
pairs so that the task set matches our end-to-end coverage objective;
(3)~\emph{RD-Probe*}, which adapts RD-Probe~\cite{ding2024rdprobe} by preserving
the two-phase structure: Phase-I randomized coverage followed by Phase-II
per-interface deficit-driven repair approximated via hash-based path
resolution; (4)~\emph{D-Pingmesh (Cold Start)}, which runs
Algorithm~\ref{alg:probe-generator} from scratch in each slot; and
(5)~\emph{D-Pingmesh}, which runs the incremental update of
Algorithm~\ref{alg:incremental-update}, reusing and repairing the prior
pinglist. deTector* and RD-Probe* run under the same accepted-path model as the
other generators.

\myparagraph{Implementation.} D-Pingmesh's probe generator with incremental
update is implemented in about 3.9K lines of Python.
The Rho-EM localizer is implemented in about
2.4K lines of Python. We use ns-3~\cite{ns3} to replay TPE, TE, and probe tasks for packet-level simulation.
The ns-3 replay implementation is about 3.3K lines of C++. All generation
and localization algorithms, as well as the ns-3 simulation, run on a
machine with Ubuntu 22.04.5 LTS, a server-grade CPU at 2.60~GHz, and
256~GiB of memory. 
\rebuttalrev{
We also deploy a physical testbed comprising one 192$\times$192 OCS and four pods, each containing a datacenter switch and an x86 server (Section~\ref{subsec:physical-testbed}).
}

\subsection{Main Generator Comparison}
\label{subsec:generator}

\myparagraph{Overall comparison.}
\bugfixrev{
Figure~\ref{fig:main-five-algo} compares the five algorithms under their default configurations across 96 slots of each F5, F12, and F20 trace.
All five algorithms reach near-perfect end-to-end coverage, with trace-level averages ranging from 99.95\% to 100\%.
They differ in probe volume and generation time.
Pingmesh generates roughly 712K tasks per slot on average, while deTector* and RD-Probe* reduce this to roughly 16.4K and 29.8K, respectively.
D-Pingmesh maintains this near-perfect coverage with an average of 10,036 tasks per slot.
Pingmesh, RD-Probe*, and deTector* use 71.0$\times$, 2.97$\times$, and 1.63$\times$ as many tasks as D-Pingmesh, respectively.
RD-Probe* and deTector* require an average of about 10~s and 208~s per slot, respectively, while D-Pingmesh averages only 2.462~s.
Averaged across all 22 traces, D-Pingmesh achieves 99.99\% coverage using 10.11K tasks and 2.50~s of generation time per slot.
}

\begin{figure}[t]
\centering
\begin{subfigure}[t]{\linewidth}
\centering
\includegraphics[width=0.90\linewidth]{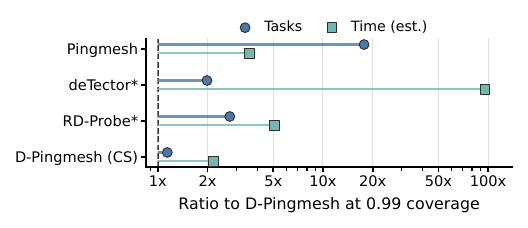}
\caption{Same 0.99 coverage}
\end{subfigure}
\begin{subfigure}[t]{\linewidth}
\centering
\includegraphics[width=0.90\linewidth]{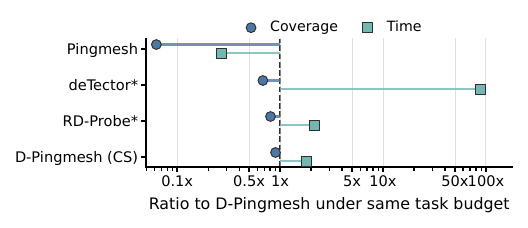}
\caption{Same task budget}
\vspace{-1ex}
\end{subfigure}
\caption{Controlled probe-generation comparisons, normalized to D-Pingmesh.}
\Description{Two compact ratio plots normalized to D-Pingmesh. The top plot
compares task and time ratios needed to reach 0.99 coverage. The bottom plot
compares coverage and time ratios under the same per-slot D-Pingmesh task
budget.}
\label{fig:controlled-probe-generation}
\end{figure}

\myparagraph{Controlled comparisons.}
We isolate coverage target
and task budget by running two experiments on the F20 trace
(Figure~\ref{fig:controlled-probe-generation}), normalizing every ratio to
D-Pingmesh: a same-coverage experiment where each generator \bugfixrev{reaches at least} 0.99
end-to-end coverage, and a same-budget experiment where each generator is capped
at D-Pingmesh's per-slot task count.
\bugfixrev{For Figure~\ref{fig:controlled-probe-generation}(a), we estimate and subtract
each method's coverage-checking overhead from its generation time.}
Under the 0.99 coverage target, deTector* is \bugfixrev{95.7}$\times$ slower, exposing the computational cost of PMC-style greedy
selection. Pingmesh needs \bugfixrev{17.7}$\times$ more tasks, confirming that
uniform server-pair probing is fundamentally mismatched with reconfigurable path
structures. RD-Probe* needs \bugfixrev{2.73}$\times$ more tasks and \bugfixrev{5.11}$\times$ more time.
Under the same task budget, Pingmesh takes \bugfixrev{0.269}$\times$ the time but reaches
only \bugfixrev{0.063}$\times$ of D-Pingmesh's coverage; deTector* and RD-Probe* reach
\bugfixrev{0.681}$\times$ and \bugfixrev{0.807}$\times$ coverage.

\myparagraph{Incremental-update ablation.}
To demonstrate the effectiveness of incremental update, we compare D-Pingmesh
with a Cold Start variant that disables incremental update and regenerates
probes from scratch in each slot.
\bugfixrev{
In the main comparison, incremental update maintains near-perfect coverage, reduces probe task volume, and achieves a 2.37$\times$ speedup, cutting average generation time from 5.827~s/slot to 2.462~s/slot (Figure~\ref{fig:main-five-algo}).
}
The controlled experiments confirm this advantage
(Figure~\ref{fig:controlled-probe-generation}): Cold Start needs \bugfixrev{an estimated 2.16}$\times$
more time under the same coverage target and \bugfixrev{1.82}$\times$ more time under the
same task budget.

\begin{figure}[t]
\centering
\includegraphics[width=\linewidth]{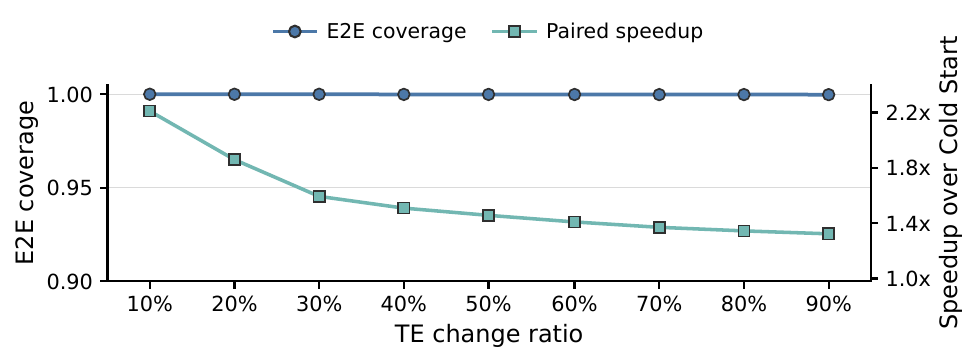}
\caption{D-Pingmesh robustness under controlled TE changes.}
\Description{A dual-axis line plot shows end-to-end coverage and incremental generation speedup relative to Cold Start as the TE change ratio increases from 10 percent to 90 percent.}
\label{fig:te-change-robustness}
\end{figure}

\myparagraph{Robustness to TE changes.}
We evaluate D-Pingmesh under controlled TE changes on F20. Each slot keeps the
TPE and TE next-hop/interface set fixed while we randomly vary the weights of a
fraction of \bugfixrev{eligible TE groups between consecutive slots}, from 10\% to 90\%
\bugfixrev{over 95 updates per setting, excluding the initial slot}
(Figure~\ref{fig:te-change-robustness}).
\bugfixrev{The per-group weight-change magnitude is calibrated to the average change observed in the F20 trace with unchanged next-hop sets.}
Coverage remains near 100\%
throughout. Generation time grows gradually, \bugfixrev{but compared with Cold Start on the same TE states, incremental update is 2.21$\times$ faster at 10\% group change and remains 1.32$\times$ faster even when 90\% of groups change weight.}

\begin{figure}[t]
\centering
\includegraphics[width=0.98\linewidth]{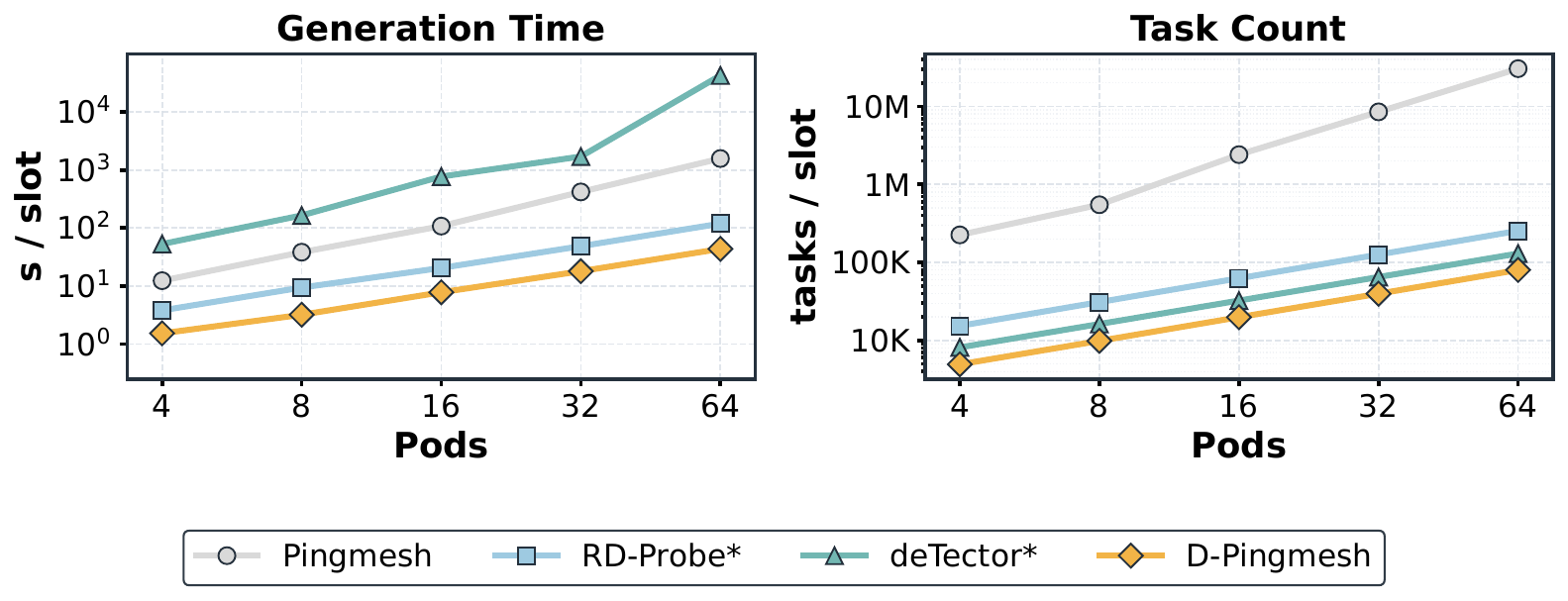}
\caption{Topology scalability across 4--64 pods.}
\Description{Two line charts show generation time per slot and tasks per slot
as the number of pods increases from 4 to 64. D-Pingmesh grows slowly in both
time and tasks, RD-Probe grows monotonically at a moderate rate, Pingmesh grows
rapidly in task count, and deTector grows rapidly in generation time.}
\label{fig:topology-scalability}
\end{figure}

\myparagraph{Topology scalability.}
We further evaluate the scalability of D-Pingmesh by varying the number of pods
from 4 to 64 on the F20 trace.
\bugfixrev{For D-Pingmesh, each scale uses 12 distinct Gemini TE snapshots over a
demand-weighted TPE; the plotted values average the 10 update slots 2--11.
The Pingmesh, RD-Probe*, and deTector* curves retain their original control-plane
inputs and results. D-Pingmesh achieves above 99\% coverage in every measured
slot at every scale}, so
Figure~\ref{fig:topology-scalability} focuses on generation time and task count.
D-Pingmesh scales from \bugfixrev{4.9K} tasks and \bugfixrev{1.54}~s/slot at 4 pods
to \bugfixrev{79.9K} tasks and \bugfixrev{43.78}~s/slot at 64 pods.
RD-Probe* also scales smoothly, growing from 15.4K tasks and
3.82~s/slot to 254.0K tasks and 119.2~s/slot. Pingmesh expands
to 30.5M tasks/slot at 64 pods, and deTector* reaches 42,510~s/slot.
Both become impractical at this scale.
\rebuttalrev{Assuming continuous probing with one 64-byte aggregate
record per task every 30 seconds, the 64-pod lists imply 5.62~TB/day for Pingmesh versus
14.73~GB/day for D-Pingmesh.}

\myparagraph{Incremental generation across TPE transitions.}
\rebuttalrev{
We evaluate 21 independently initialized transitions from F1 to F2 through F21 to F22 on the 8-pod topology.
Each transition updates the source trace's slot-95 pinglist under the next trace's TPE and slot-0 TE snapshots, changing both OCS connectivity and routing. 
Across these controlled transitions, incremental generation achieves mean end-to-end $k=4$ coverage of 99.97\%, with 10,620 tasks and 3.995~s per update on average.
The results demonstrate that D-Pingmesh remains effective for TPE changes.
}

\subsection{Fault Localization}
\label{subsec:fault-loc}

We use ns-3 packet-level simulation to evaluate D-Pingmesh's fault localization algorithm Rho-EM.
\bugfixrev{We use the same 8-pod topology and TPE/TE snapshots as the generation experiments.
We evaluate on the F20 trace because it exhibits substantial TE-boundary churn, using a fixed D-Pingmesh probe list for the localization experiments.}
Faults are injected on randomly selected ToR and Agg switch interfaces.

Each probe task executes once per second over a 30-second window. The localizer
uses the aggregate per-task attempt and failure counts at the end of each
window. We compare two scenarios: in the stable-TE scenario, the TE snapshot
remains fixed throughout the window; in the boundary scenario, the TE state
transitions from the old TE snapshot to the new one in four equal batches of
pods, so the aggregate loss count for each task reflects a mixture of
measurements taken under the old and new path hypotheses.

We compare four localization variants.
\emph{Stable TE} and \emph{Boundary TE + New Path} are simplified variants of
Rho-EM with the per-task $\rho$ fixed to 1.0 for the new-TE path. Stable TE
runs under the stable-TE scenario, where this single-path assumption holds.
Boundary TE + New Path runs under the boundary scenario, where fixing $\rho$ to
1.0 discards old-path evidence.
\emph{Boundary TE + deTector PLL} applies deTector's PLL
algorithm~\cite{peng2017detector} on the
same new-TE path view. \emph{Boundary TE + Rho-EM} runs under the boundary scenario and applies the
Rho-EM algorithm to estimate each task's unknown old/new path mixture.
We evaluate under two fault configurations, each repeated over ten random seeds
for fault placement. In the first, we vary the number of faulty links over
$\{1,2,5,10,20,30\}$ at 10\% loss. In the second, we vary the loss rate over
$\{5,10,20,30,50,100\}\%$ with 10 faulty links.

\begin{figure}[t]
\centering
\includegraphics[width=\linewidth]{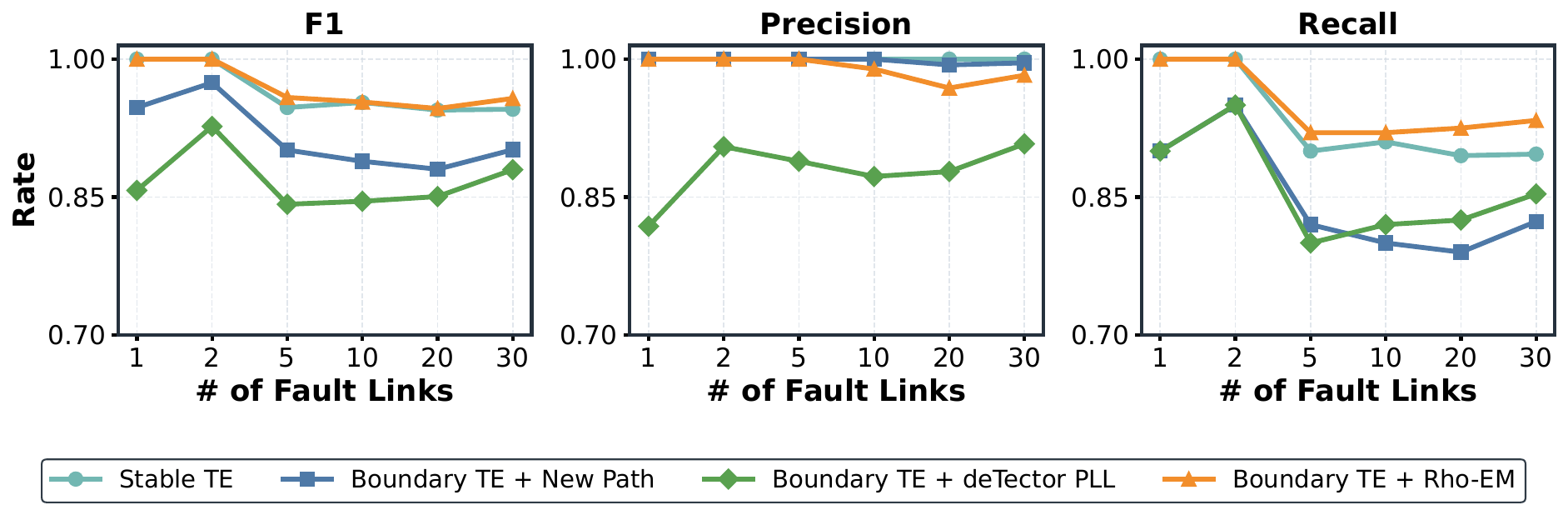}
\caption{Fault localization with varying number of faulty links.}
\Description{Three line plots show F1, precision, and recall versus the number
of fault links for Stable TE, Boundary TE with new paths, Boundary TE with
deTector PLL, and Boundary TE with Rho-EM.}
\vspace{-1ex}
\label{fig:faultloc-rho-f20-fault-count}
\end{figure}

\begin{figure}[t]
\centering
\includegraphics[width=\linewidth]{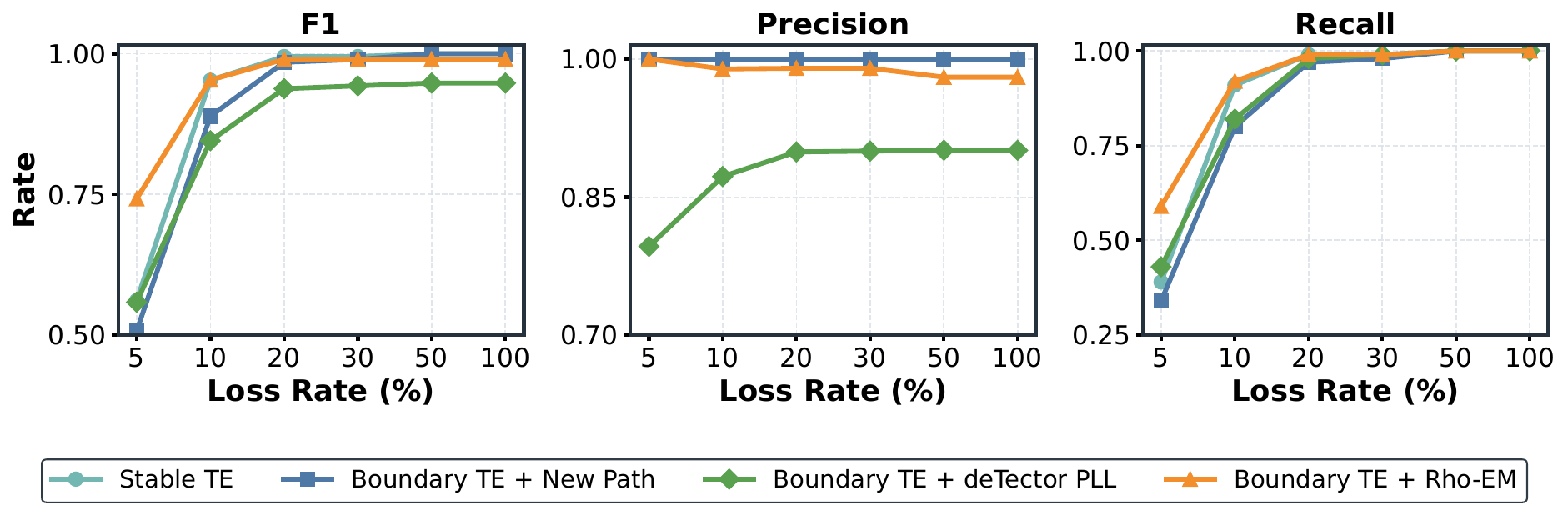}
\caption{Fault localization with varying loss rate.}
\Description{Three line plots show F1, precision, and recall versus link-loss
rate for Stable TE, Boundary TE with new paths, Boundary TE with deTector PLL,
and Boundary TE with Rho-EM.}
\vspace{-1ex}
\label{fig:faultloc-rho-f20-loss-rate}
\end{figure}

\revised{Figures~\ref{fig:faultloc-rho-f20-fault-count} and~\ref{fig:faultloc-rho-f20-loss-rate}
show localization results under varying fault counts and loss rates.
Across all 120 cases, Rho-EM achieves 0.953 micro F1, compared with 0.871
for deTector PLL and 0.907 for New Path. Across both sweeps, Rho-EM maintains
high precision and achieves higher recall than the boundary single-path baselines.
The advantage is more pronounced at low loss rates, where weak observations
amplify boundary ambiguity.}

\rebuttalrev{
We further examine Rho-EM's sensitivity to the iteration cap.
Across all 120 cases, micro F1 increases from 0.946 with $M=0$ to 0.953 with $M=1$.
Modeling both paths with fixed mixture weights therefore already captures most of the average gain over the single-path baseline.
At 5\% per-link loss with 10 faulty links, increasing the iteration cap from $M=0$ to $M=1$ yields a larger micro F1 improvement, from 0.630 to 0.742.
Over the full evaluation grid, mean localization time is 1.070~s per case for
$M=0$ and 1.235~s per case for $M=1$.
Larger iteration caps yield no further micro F1 improvement in this evaluation.
}

\begin{figure}[t]
\centering
\begin{subfigure}[t]{\linewidth}
\centering
\includegraphics[width=\linewidth]{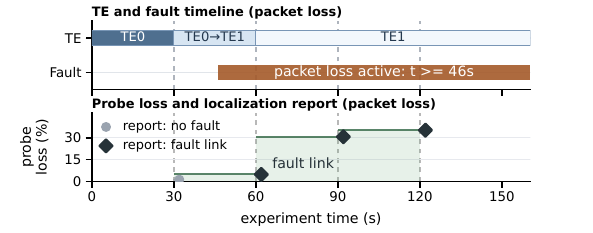}
\caption{Persistent packet-loss fault}
\end{subfigure}
\begin{subfigure}[t]{\linewidth}
\centering
\includegraphics[width=\linewidth]{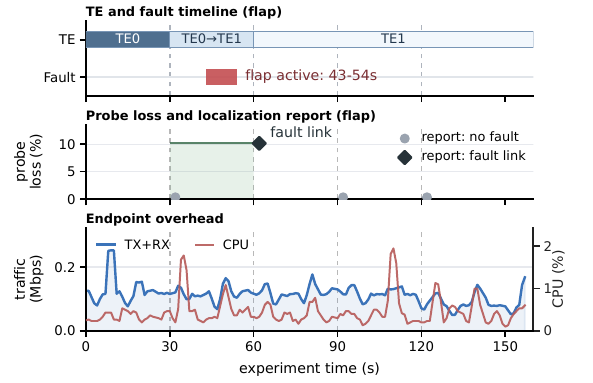}
\caption{Transient link flap fault}
\end{subfigure}
\caption{Physical testbed validation with 30-second observation windows.} \vspace{-1ex}
\label{fig:testbed-validation}
\end{figure}

\subsection{Physical Testbed Validation}\label{subsec:physical-testbed}

We validate D-Pingmesh on a physical testbed with one 192$\times$192 OCS and
four minimal pods, each consisting of a datacenter switch and an x86 server. Each run uses 30-second observation windows, with a TE transition in the middle window, 30--60s. Each probe is a UDP echo packet with a per-task five-tuple and DSCP tag.
Figure~\ref{fig:testbed-validation} shows the results for two fault scenarios: persistent packet loss and transient link flap.

\myparagraph{Persistent packet-loss.} Packet loss is emulated by MQC CAR rate limiting on the switch, combined with business traffic that shares the same DSCP-marked behaviors as the probes. Before injection, D-Pingmesh reports no fault.
Once the fault becomes active during the TE transition, D-Pingmesh localizes it to the injected inter-pod link in that same window, despite only 4.8\% measured probe loss, and continues to report the fault in subsequent windows.

\myparagraph{Transient link flap.} The link flap is emulated by inserting and then removing MQC deny rules on the switch that match the probe's five-tuple and DSCP. The link flap is active only during
the TE-transition window. D-Pingmesh localizes the fault to the injected inter-pod link in that window
with 10.1\% measured probe loss, and returns to no-fault reports in the
following stable-TE windows.

\myparagraph{Time to localization.} D-Pingmesh provides second-scale fault localization. In both scenarios, the fault becomes effective roughly 15 seconds into the TE-boundary window and is localized at the window boundary, with the result available within 2 seconds, dominated by result collection and aggregation. Shorter observation windows can further reduce fault detection latency.  

\myparagraph{Host overhead.} Host overhead is small. During the experiment,
average CPU utilization is 0.43\% and average per-server TX+RX traffic is
0.10~Mbps. 
The testbed runs 34 probes per round across four servers, or 8.5 per server on average, whereas the main 8-pod experiment averages 10,036 tasks across 2,048 servers, or about 4.9 per server.

\section{Related Work}

\myparagraph{Datacenter active probing and diagnosis.}
Pingmesh~\cite{guo2015pingmesh} established production-scale active probing for
datacenter monitoring. R-Pingmesh~\cite{liu2024rpingmesh} extended this to RoCE
networks with service-aware probing and in-network drop differentiation. For fault diagnosis,
007~\cite{arzani2018zerozeroseven} aggregates end-host observations to localize
packet-drop faults,
NetBouncer~\cite{tan2019netbouncer} uses IP-in-IP bouncing to probe
designated paths, and ByteTracker~\cite{guo2025bytetracker} tracks probe paths with
centralized probers and switch mirroring. deTector~\cite{peng2017detector} selects topology-aware probe matrices for joint
coverage and loss localization, and
RD-Probe~\cite{ding2024rdprobe} combines randomized and deterministic generation
to guarantee L2 coverage at scale. 
\revised{These earlier systems assume stable forwarding paths.
OmniPath Ping~\cite{yang2026omnipath} uses switch-host co-design for service
tracing under packet spraying, whereas D-Pingmesh targets path changes across
TPE/TE reconfigurations.}

\myparagraph{Runtime forwarding verification.}
\rebuttalrev{
VeriDP~\cite{zhang2016mind} monitors consistency between control-plane policies
and actual packet forwarding, while Hydra~\cite{renganathan2023hydra} verifies
packet processing against specifications at runtime.
Both complement D-Pingmesh's probing and fault localization.
}

\myparagraph{Traffic engineering.}
Recent work on precise datacenter
TE~\cite{chen2024precise} studied realizing desired weight distributions under
constrained hardware resources.
Gemini~\cite{zhang2021gemini} jointly optimizes inter-pod topology and routing
for reconfigurable datacenter networks.
FIGRET~\cite{liu2024figret} enhances TE robustness against traffic bursts with
fine-grained per-source-destination-pair protection.
SSDO~\cite{mao2026ssdo} accelerates large-scale TE via a sequential
solver-free algorithm. 
We use TE output as the forwarding state for probe generation and
fault localization.

\myparagraph{Reconfigurable datacenter networks.}
\revised{OSA~\cite{chen2012osa} established the case for optical reconfiguration,
while Sirius~\cite{ballani2020sirius} demonstrated a flat fabric with nanosecond switching.
Jupiter Evolving~\cite{poutievski2022jupiter} and Lightwave~\cite{liu2023lightwave}
demonstrated production OCS fabrics.
TPUv4~\cite{jouppi2023tpu} deployed OCS to reconfigure machine-learning interconnects,
while InfiniteHBD~\cite{shou2025infinitehbd} uses OCS transceivers to build high-bandwidth domains for LLM training.
TROD~\cite{cao2023trod,cao2021trod} co-designs threshold-based routing and topology for optical datacenters.
NegotiaToR~\cite{liang2024negotiator} uses distributed on-demand scheduling.
OpenOptics~\cite{lei2026openoptics} provides an open framework for optical architectures,
and SyncWise~\cite{lei2026syncwise} studies time synchronization for reconfigurable networks.}

\section{Discussion}

\myparagraph{Generality.}
D-Pingmesh formulates probe generation and fault localization over the
abstractions of topology, TE weights, and WCMP hash behavior, without relying on
properties specific to OCS hardware. The approach applies to any network in which
forwarding paths change on timescales shorter than the measurement window,
including Low Earth Orbit (LEO) satellite constellations, where inter-satellite links
reconfigure under orbital motion~\cite{lai2025leocc,li2025small}. 

\myparagraph{Fault model and observation granularity.}
Rho-EM's likelihood model assumes per-interface independent loss with a stable
rate within one probe window.
The model does not capture failures with shared-risk structure, such as an
OCS port affecting multiple circuits, nor time-varying loss rates.
\rebuttalrev{Task-level $\rho_q$ accommodates heterogeneous old/new path mixtures
using the existing aggregate counters. A per-pod commit-time model uses fewer
parameters, but ties the mixtures to a shared transition time within each pod.
Sub-window reporting divides a probe window into shorter intervals and reports
per-task attempt and failure counts for each interval. This provides finer
temporal evidence, but increases reporting volume and reduces samples per
interval at a fixed probing rate.}

\myparagraph{\rebuttalrev{Deployment trade-off.}}
\rebuttalrev{D-Pingmesh preserves deployment simplicity: agents execute
pinglists and report aggregate counters, while switches remain unchanged.
Additional computation and state reside at the controller.
D-Pingmesh suits large, frequently reconfigured fabrics; Pingmesh-style
monitoring remains simpler for small, mostly static networks when probing
and reporting costs are acceptable.}

\myparagraph{Towards faster TE cadences.}
\rebuttalrev{Our evaluation uses 15-minute TE windows~\cite{mao2026ssdo}.
At the 8-pod scale, D-Pingmesh averages 2.462~s/slot for probe generation (Figure~\ref{fig:main-five-algo}).
Faster TE cadences may require shortening the 30-second observation window,
reducing probe attempts per interface.
Raising the per-interface coverage target $k$ could help compensate,
at the cost of additional probing.}

\myparagraph{\rebuttalrev{More frequent TPE reconfiguration.}}
\rebuttalrev{If TPE reconfiguration moves to hourly or minute-level operation,
changes in optical connectivity would create more frequent pinglist invalidation
and boundary ambiguity, increasing the importance of Challenges 2 and 3.
Our controlled TPE-transition experiment tests incremental generation after the
new state is available; handling observations during optical switching requires
further evaluation.}

\section{Conclusion}

D-Pingmesh co-designs probing and fault localization with the TPE/TE control
plane for reconfigurable datacenter networks. It combines probability-guided
probe generation, incremental pinglist updates, and Rho-EM, which jointly
refines per-task old/new-path mixtures and the fault set to resolve boundary
ambiguity. Compared with prior systems, D-Pingmesh achieves equivalent coverage
with fewer tasks and lower generation time, while improving fault localization
under continuous reconfiguration.

\begin{acks}
We thank our shepherd, Cheng Tan, and the anonymous reviewers for their helpful feedback. This work was supported by the National Natural Science Foundation of China (NSFC) under Grants 62132009, 62221003, and 62602347.
\end{acks}

\bibliographystyle{ACM-Reference-Format}
\begingroup
\interlinepenalty=6000
\bibliography{paper}
\endgroup

\end{document}